\documentclass[12pt]{article} 
\pdfoutput=1
\usepackage{amsmath,color,epsfig,amsfonts,fullpage, natbib,rotating,
  subfigure, algorithmic, algorithm, tabularx} 
  \usepackage{setspace}
  \usepackage{mathrsfs}
\usepackage[normalem]{ulem} 
\usepackage[colorlinks=true, pdfstartview=FitV, linkcolor=blue,
citecolor=blue, urlcolor=blue]{hyperref}
\usepackage{verbatim} % for comments environment
\usepackage{ctable} % table.  
\definecolor{britishracinggreen}{rgb}{0.0, 0.26, 0.15}
\definecolor{burgundy}{rgb}{0.5, 0.0, 0.13}

\newcommand{\nch}{\left\{\begin{array}{ll}}\newcommand{\ech}{\end{array}\right.}

\newcommand{\vv}[1]{\mbox{\boldmath $#1$}}
\DeclareMathOperator*{\argmax}{arg\,max}

\newcommand*{\manuscriptref}[1]{(M\ref{#1})}
\newcommand*{\manuscriptrefnp}[1]{M\ref{#1}}

\newtheorem{prop}{Proposition}
\newtheorem{definition}{Definition}

\newtheorem{innercustomprop}{Proposition}
\newenvironment{customprop}[1]
{\renewcommand\theinnercustomprop{#1}\innercustomprop}
{\endinnercustomprop}

  \newtheorem{lemma}{Lemma}

\begin{document}

%
%
%%%%%%%%%%%%%%%%%%%%%%%%%%%%%%%%%%%%%%%%%%%%%%%%%%%%%%%%%%%%%%%%%%%%%%%%%%%%%%%
%
%\if1\blind
%{
%  \title{\bf Anchored Bayesian Gaussian Mixture Models}
%  \author{Deborah Kunkel \thanks{ This work was partially supported by the National Science Foundation under Grant
%No. SES-1424481.
%    The authors gratefully acknowledge Steven MacEachern for his helpful comments.}\hspace{.2cm}\\
%    Department of YYY, University of XXX\\
%    and \\
%    Mario Peruggia \\
%    Department of Statistics, The Ohio State University}
%  \maketitle
%} \fi
%
%\if0\blind
%{
%  \bigskip
%  \bigskip
%  \bigskip
%  \begin{center}
%    {\LARGE\bf Title}
%\end{center}
%  \medskip
%} \fi
%
%\medskip
\begin{center}
\section*{Anchored Bayesian Gaussian Mixture Models}
\vspace*{-3mm}

\textbf{Deborah Kunkel$^1$ and Mario Peruggia$^2$}

\vspace{.1cm}
1. School of Mathematical and Statistical Sciences,
Clemson University,
Clemson, SC, USA\\
2. Department of Statistics,
The Ohio State University,
Columbus, OH, USA\\
\end{center}

\vspace*{-8mm}
\begin{abstract}
\vspace*{-3mm}
\noindent
Finite mixtures are a flexible modeling tool for irregularly shaped
densities and samples from heterogeneous populations.  When modeling
with mixtures using an exchangeable prior on the component features,
the component labels are arbitrary and are indistinguishable in
posterior analysis.  This makes it impossible to attribute any
meaningful interpretation to the marginal posterior distributions of
the component features. We propose a model in which a small number of
observations are assumed to arise from some of the labeled component
densities.  The resulting model is not exchangeable, allowing
inference on the component features without post-processing.  Our
method assigns meaning to the component labels at the modeling stage
and can be justified as a data-dependent informative prior on the
labelings.  We show that our method produces interpretable results,
often (but not always) similar to those resulting from relabeling
algorithms, with the added benefit that the marginal inferences
originate directly from a well specified probability model rather than
a {\em post hoc\/} manipulation. We provide asymptotic results leading
to practical guidelines for model selection that are motivated by
maximizing prior information about the class labels and
\mbox{demonstrate our method on real and simulated data.}
\end{abstract}

\clearpage
\vspace*{-13mm}
\section{Introduction}
\vspace*{-5mm}

Finite mixture models are flexible tools that are often applied to
data from heterogeneous populations or from distributions with
irregularly-shaped densities. In these context, they can produce
useful approximations to the unknown density functions in both
univariate and multivariate settings \citep{Fru2006,
  marin2014bayesian, rossi2014}.  Results concerning the accuracy and
consistency of the approximations (as the number of components
increases at an appropriate rate) have been established both in the
frequentist and in the Bayesian settings \citep{RoederWasserman1997,
  genovese2000, Norets2012}.  In many situations, if the true density
is well-behaved in the tails, satisfactory approximations can be
obtained using a small or moderate number of mixture components.

When mixture distributions are used to model heterogeneous
populations, the mixture components are thought to represent clusters
of similar units.  Such analyses are found in areas such as medicine,
the social sciences, and genetics, where identifying subgroups of
similar individuals may help to generate hypotheses for future
research. When subgroup identification is an important goal, as it is
in this paper, the parameters of the component distributions provide
population-level information about the features of groups, which can
elucidate the overarching patterns of heterogeneity within a
population.  Therefore, accurate estimates of component-specific
parameters, with attendant measures of uncertainty, become a vital
element of inference.  Further, the mixture model allows estimation of
a probabilistic clustering structure from the data.
  
Adopting standard notation \cite[e.g.,][]{Fru2006},
we represent the likelihood for a $k$-component finite
mixture model for a response \mbox{$\vv{y}=(y_1,\ldots,y_n)$ as}
\vspace*{-5mm}
\begin{align}\label{GMM}
  f( \vv{y}|\vv{\gamma},\vv{\eta}) &= \prod_{i=1}^n \sum_{j=1}^k\eta_j
                                     p(y_i|\gamma_j), 
\end{align}
where $p(\cdot|\gamma_j)$ denotes the $j$th component density.
%the vector 
The model is parameterized by
$\vv{\eta}=(\eta_1,\ldots,\eta_k)$, 
%contains 
 the vector of 
mixture proportions whose elements sum to one, and $\vv{\gamma} =(\gamma_1,\ldots,\gamma_k)$, the vector of features of the component densities.  It is often helpful to write the
model (\ref{GMM}) hierarchically, using latent variables
$\vv{S} = \{ S_1,\ldots,S_n\}$,
$S_i \in \{ 1,\ldots,k\}$, $i=1,\ldots,n$, to indicate component
membership.  
The resulting likelihood is
\vspace*{-3mm}
\begin{align}\label{GMM_latentallocations}
  % f(y_i|S_i=s_i,\vv{\gamma}) &= \phi(y_i;\gamma_{s_i}), \;\;\; P(S_i
  % = j|\vv{\eta}) = \eta_j, \quad i=1,\ldots,n. \\
  f(\vv{y}|\vv{s},\vv{\gamma}) &= \prod_{i=1}^np(y_i|\gamma_{s_i}), \quad \mbox{where } \;\;\; P(S_i = j|\vv{\eta}) = \eta_j, \quad j=1,\ldots, k; \quad i=1,\ldots,n.
\end{align} 

\vspace{-3mm} \noindent If the population comprises well-understood
groups, it is appropriate to incorporate information regarding the
groups' relative locations and scales into the prior on
$(\vv{\gamma}, \vv{\eta})$.  Often, however, little is known about
these groups ahead of time, and it is natural to assume prior
exchangeability of component features. Let $q$ index the $k!$
  possible permutations of the integers $1,\ldots,k$
  and let
  $\rho_q(\cdot)$ relabel its argument according to the $q$th
  permutation. We will use the same symbol $\rho_q(\cdot)$ to
  denote the action of the $q$th permutation on a given index between 1 and
  $k$ and on the elements of a vector argument.
  For example, if the $q$th permutation of $(1,2,3,4)$ is
  $(1,3,4,2)$, then $\rho_q(3)=4$ and
$\rho_q(a_1,a_2,a_3,a_4)=(a_1,a_3,a_4,a_2)$. 

An exchangeable prior with density $\pi$ satisfies
$\pi(\vv{\gamma}, \vv{\eta}) = \pi(\rho_q(\vv{\gamma},\vv{\eta})),$
$q=1,\ldots,k!$.   This
specification produces a posterior distribution that inherits the same
label invariance; that is, letting $p_E(\cdot|\vv{y})$ denote the
posterior density of $(\vv{\gamma},\vv{\eta})$ under the exchangeable
model,
\begin{align}
  p_E(\vv{\gamma},\vv{\eta}|\vv{y})
  &=
    p_E(\rho_q(\vv{\gamma},\vv{\eta})|\vv{y}),
    \quad q=1,\ldots,k!.  
\end{align}
The posterior density is symmetric with respect to the $k!$ labelings
	of the components, often producing $k!$ modal regions in the
        parameter space.  
  The marginal
distributions of the component-specific parameters are identical.
When Markov Chain Monte Carlo methods are used to sample from the
posterior distribution of $(\vv{\gamma},\vv{\eta})$, a well-mixed
chain will jump from one possible labeling to another, a phenomenon
referred to as ``label-switching.''

The posterior symmetry and accompanying label-switching in no way hinders the model's predictive performance, nor does it preclude meaningful inference on objects that do not depend on the component labels.  When label-switching occurs, however,
	ergodic averages cannot be used for inference on the
	component-specific features, making it impossible to use labeled parameters to learn about distinctions among the mixture components.    Much work has been devoted to either
preventing or reversing label-switching by placing prior
constraints on the parameter space or by post-processing posterior
samples in a way that allows only one possible labeling of the mixture
components.  These approaches, particularly the post-processing
approach, are popular in practice.

Prior identifiability constraints create a non-exchangeable prior by
requiring $(\vv{\gamma},\vv{\eta})$ to lie in some sub-region of the parameter
space that is compatible with only one possible labeling.  For
example, one could require that $\eta_1 < \ldots < \eta_k$ with
probability $1$, or establish similar ordering constraints among the component feature parameters.   The limitations of these approaches are
addressed in detail by, among others, \cite{celeux} and
\cite{jasra2005}.  They are often considered too informative in their
strict restrictions of the parameter space and may not effectively
isolate a single modal region of the posterior density. It is not
always obvious, a priori, what choice of constraint is appropriate for
a problem.

Relabeling algorithms, such as those presented by
\cite{stephens2000,celeux,marin2005,ECR,bardenet1, rodriguez}, and
\cite{LiFan2016}, tend to be preferred. These algorithms specify a
loss function and find the labeling that minimizes the loss function
for each posterior sample of $(\vv{\gamma},\vv{\eta})$ and, if
sampled, $\vv{s}$.  Upon convergence, each
unique value of $(\vv{\gamma},\vv{\eta})$ is restricted to only one possible labeling. For this reason,
\cite{jasra2005} have described this strategy as a way of
automatically applying an identifiability constraint.  Relabeling
algorithms often appear to perform ``better'' than prior constraints, in that they
produce relabeled posterior samples that have
unimodal and well-separated marginal densities.  In contrast to
methods based on prior constraints, however, it is not straightforward to
obtain expressions for the joint or marginal distributions of the
elements of $(\vv{\eta},\vv{\gamma})$ corresponding to a relabeling method.  The
constrained region of the parameter space is the solution to the
iterative minimization of the chosen loss function, and, as such,
cannot be described concisely as a component of the probability model.
Because its constraints are not the result of a clearly defined prior
specification, it is difficult to evaluate
rigorously the underlying structure that the relabeling
algorithm imposes upon a problem.
It is not obvious whether
this approach can be justified as a basis for making inferential
claims about the posterior distribution of the component-specific
parameters.

We introduce a modification to the
standard finite mixture model, the {\em anchor\/}
model, in which a small number of observations are assumed to be drawn from known component densities.  This breaks the
model's label invariance in a data-dependent manner while avoiding the
strong, subjective restrictions imposed by prior identifiability
constraints.  The anchor model provides an appropriate basis upon
which to interpret the component-specific feature parameters by
assigning meaning a priori to their labels.  The proposed modeling
framework requires a modest amount of pre-processing to identify the
pre-classified observations but avoids the computational burden of
post-processing.

The rest of the paper is organized as follows. In
Section~\ref{section::anchors} we describe our
  proposed anchored mixture model and its basic
properties.   We also examine the implications of its asymptotic behavior on
the identifiability of the component-specific feature parameters. 
In Section~\ref{section::model_specification}
we outline practical strategies for
model specification.  In Sections~\ref{section::Examples}
and~\ref{section::Falldata} we present data analysis
examples that make use of our proposed methodology.
In Section~\ref{section:discussion} we present concluding
  remarks and discuss directions for possible future developments.

\vspace*{-8mm}
\section{Anchor models} \label{section::anchors} \vspace*{-5mm} The
idea of assuming known labels for some observations has been
considered by \cite{Chungetal2004}, who present this as a way of
specifying an informative prior, and, more recently, by
\cite{Egidietal2016}, who propose a post-processing strategy
that assigns labels to observations with zero \emph{posterior}
probability of being allocated to the same mixture
component.  Related approaches that disallow specific allocations of
the observations to the various mixture components have been suggested
as a means of guaranteeing propriety of the posterior distribution if
improper priors are specified \citep{DieboltRobert1994,wasserman2000}.
We build on these ideas by formalizing this strategy as a modeling
procedure that requires no post-processing of an MCMC
  sample.
%\remove{neither \deborah{post-processing} nor an MCMC  sample from
%the posterior distribution of the full exchangeable model}.
A careful assignment of a small number of
observations to specific components yields a well-defined mixture model whose components
can accurately reflect homogeneous subgroups in the population.  We
define the anchor model and describe several of its basic properties
in the following sections.

\vspace*{-5mm}
\subsection{Definition of an anchor model} 
\vspace*{-3mm}  The anchor model
modifies the finite mixture likelihood by selecting a small number of observations whose component labels are assumed to be known.   These observations will be
called \textit{anchor points}.  The resulting model can be
fully described using $k$ index sets $A_j,\,\, j=1,\ldots,k$, where $A_j$
contains the indices of those observations in the dataset that are to
be ``anchored'' to the $j$th component and $A =\{A_1,\ldots,A_k \} $
is the set of indices of all anchor points. The likelihood in~\eqref{GMM} is replaced by 
\begin{singlespace}
	\begin{align} \label{anchor_likelihood}
	\begin{split}
	f_A(y_i|\vv{\gamma},\vv{\eta}) &=\begin{cases}
	\sum_{j=1}^k\eta_jp(y_i|\gamma_j), \quad &i \notin A, \\
	p(y_i|\gamma_j), \quad &i \in A_j, \quad j=1,\ldots,k.
	\end{cases}
	\end{split}
	\end{align}
\end{singlespace}
 \noindent We use $m_j$ to denote
the number of points anchored to the $j$th component and
$m=\sum_{j=1}^{k}m_j$ to denote the total number of anchor points.
Some of the $A_j$ may be empty and the number of components that
contain one or more anchor points is denoted by $k_0 \leq k$. 
The anchor model can equivalently be represented using latent allocations: if $i$ is the index of an anchored
observation, $P(S_i = j) = 1$ for one prespecified component $j$. This
restricts the support of $\vv{S}$ so that a subset of the possible
allocation vectors has prior probability of $0$.  The hierarchical representation of the probability density for $y_i$ under anchor model $A$ is
\vspace*{-4mm} 
% pvariate Gaussian
% eta unknown
\begin{singlespace}
\begin{align} \label{anchor_likelihood_latent_allocations}
  \hspace{-3mm}
  \begin{split}
f(y_i|S_i = s_i, \vv{\gamma}) &= p(y_i|\gamma_{s_i}),  
\, P_A(S_i = j|\vv{\eta})= \begin{cases} \eta_j, \quad &i \notin A  , \\
1, \quad &i \in A_j,  \\
0, \quad &i \in A_{j'}, \quad \mbox{with }j'\in\{1,\ldots,k\}\setminus\{{j}\}, 
\end{cases}
\end{split}
\end{align}
\end{singlespace}
 \noindent
Since an observation can be anchored to at most one component, we
require $A_j\cap A_{h} = \emptyset$ for $j=1,\ldots,k_0-1$ and $h=j+1,\ldots,k_0$.  To impose a unique labeling
on each anchor model, we will further require that $A_j \neq \emptyset$
for $j=1,\ldots,k_0$, (if any components have no anchor points, they
will be labeled $k_0+1,\ldots,k$) and that $\min_i(A_1) <
\min_i(A_2)<\ldots < \min_i(A_{k_0})$.
% for a fixed $i$ and $j' < j$, $i \in A_{j}$ only if $i' \in A_{j'}$
% for some $i' < i $. In other words, starting with observation $i=1$,
% if observation $i$ is anchored, it must belong to an observation
% should
% be %In other words, we require that $A_1$ contains the smallest
% index of an anchored observation, and observation $i$ can be anchors
% to a nonempty $A_j$ or the smallest labeled of the empty $A_j$. 
We will denote the values of
the anchor points by $\vv{x}= (\vv{x}_1,\ldots,\vv{x}_k)$, where
$\vv{x}_j = \{ y_i: i \in A_{j} \}$.  

For the remainder of this paper, we consider the anchored Gaussian mixture model, in which $p(\cdot|\vv{\gamma}_j)$ is a $p$-variate Normal distribution with density denoted by $\phi_p(\cdot;\vv{\gamma}_j)$ and $\vv{\gamma}_j=(\theta_j, \Sigma_j)$, where $\theta_j$ and $\Sigma_j$ are the mean vector and covariance matrix of the $j$th component density, respectively.
We assume that $\vv{\eta}$ and $\vv{\gamma}$ are a priori independent and that their distributions satisfy two conditions:
\vspace{-3mm}
\begin{enumerate} \label{item:conditions}
	\item[{\bf C.1:}] The prior on the mixture proportions, $\eta_1,\ldots, \eta_k$, is a Dirichlet
	distribution with concentration parameter $\alpha \vv{1}_k$, where $\vv{1_k}$ is a vector of $k$
	ones and $\alpha>0$.   \\[-30pt] 
	\item[{\bf C.2:}] The prior
	on $\vv{\gamma} = (\gamma_1, \ldots, \gamma_k)$ has the form
	$\prod_{j=1}^k\pi(\vv{\gamma}_j|\xi)$, for  some
	continuous density~$\pi$ with (possibly unknown) hyperparameters $\xi$, and $\pi$ is positive on an open subset of
	the parameter space. \\
	%\item[C.2:] The prior
	%on $\vv{\gamma}$ has a density of the form
	%\deborah{$\pi(\vv{\gamma}) = \prod_{j=1}^k\pi(\vv{\gamma}_j)$, for  some
	%density~$\pi^\prime$}.
	For ease of notation, we  will suppress $\xi$ from notation and refer to the prior density of $\vv{\gamma}_j$ using $\pi(\vv{\gamma}_j)$. 
\end{enumerate}

\subsection{Basic properties} 
\vspace*{-3mm}
In this section, we discuss some features of an anchor model
which may be readily
  understood via the latent allocation representation
in~\eqref{anchor_likelihood_latent_allocations}.  The notation $\mathscr{S}$ will denote the set of all $k^n$ possible
allocation vectors of length $n$, the sample space of the latent
variable $\vv{S}$ under the exchangeable model.  
Each allocation
vector separates the data into $k$ or fewer groups of observations and
we will refer to each unique grouping as a ``partition'' of the data.
All allocation vectors that are equal up to a relabeling of the
component labels induce the same partition of the data; e.g., we will
say that the allocations $(1,2,2,2,3)$ and $(2,1,1,1,3)$ induce the
same partition.  

Let $\mathscr{S}^A$ be the subset of allocations
that have nonzero probability under $A =\{A_1,\ldots,A_k \}$, an anchor model with $m$
anchor points. 
Then $\mathscr{S}^A$ contains $k^{n-m}$ elements and $A$ assigns
probability zero to every allocation
that satisfies $s_i = s_{i\prime}$, for some $i \in
A_j$ and $i\prime \in A_{j\prime}$, $j \neq j\prime$. Consequently, all partitions which
assign $i$ and $i\prime$ to the same group have probability zero.  
This is a key difference between anchor models and relabeling
methods that also restrict the set of allocations such as those of \cite{ECR} and \cite{rodriguez}. The relabeling
in those references creates a restricted
set of allocations that includes exactly one labeling for each
partition but does not eliminate any partitions.

Under mild conditions,
the anchor model
admits no labeling ambiguity, thus eliminating the symmetry of the
exchangeable model.   The following proposition is a direct consequence of the definition of the anchor models and is proved in the Appendix. 
%\paragraph{Proposition 2.} 
\vspace{-3mm}
\begin{prop} \label{prop:unique_labeling}
 The following two statements hold under conditions C.1 and C.2.
 \begin{enumerate}
 	\item    	An anchor model $A
 	=\{A_1,\ldots,A_k \}$ imposes a unique labeling on each partition 
 	that has nonzero probability if and only if $A_1,\ldots,A_{k-1}$ are
 	non-empty; that is, $k_0\geq k-1$. 
 	\item  For any $j\leq k_0$, $j\neq j\prime$, the marginal
 	posterior density of $\vv{\gamma}_j$ is distinct from the marginal
 	posterior density of $\vv{\gamma}_{j\prime}$ with probability~1.
 \end{enumerate}
\end{prop}
To get an intuition of why the second statement of Proposition~\ref{prop:unique_labeling} holds,
 notice that the
observations 
anchored to component $j$ contribute to the updating of the distribution of $\vv{\gamma}_j$ with probability~1, but never contribute to the updating of $\vv{\gamma}_{j\prime}$.
   Any anchor model satisfying the conditions in Proposition~\ref{prop:unique_labeling} produces distinct posterior distributions for the component-specific features.  
   The next two sections describe properties that aid in evaluating which anchor models
   are most effective at separating distinct groups in the sample.

\vspace*{-5mm}
\subsection{Model evidence} 
\vspace*{-3mm}  One key advantage of the anchor model is that
each set of anchor points results in a unique, well-defined
probability model, making it possible to compare different anchor
models using standard model selection criteria.  The goodness-of-fit
of an anchor model $A$ %with $m$ anchored points 
may be evaluated using
the model marginal likelihood, defined as
$ m_A(\vv{y}) = \int
f_A(\vv{y}|\vv{\gamma}, \vv{\eta})\pi(\vv{\gamma}) \pi(\vv{\eta})d(\vv{\gamma},\vv{\eta})$. This expression
can be expressed in terms of the latent allocations as 
\vspace{-8mm}
\begin{align} \label{anchor_marginal_likelihood}
m_A(\vv{y}) &=\sum_{\vv{s}\in \mathscr{S}^A}m(\vv{y}|\vv{s})p_A(\vv{s}),
\end{align}

\vspace{-5mm} \noindent
where we define 
$m(\vv{y}|\vv{s})= \int
f(\vv{y}|\vv{\gamma},\vv{s})\pi(\vv{\gamma})d\vv{\gamma}$ and $p_A(\vv{s}) = \int p_A(\vv{s}|\vv{\eta})\pi(\vv{\eta})d\vv{\eta}$.  Based on
Equation~\eqref{anchor_marginal_likelihood}, the goodness of fit of an
anchor model $A$ will be determined by a weighted average of the values $m(\vv{y}|\vv{s})$
over all allocations in $\mathscr{S}^A$.  The terms
$m(\vv{y}|\vv{s})$ describe the fit of the model conditional on the
partition induced by $\vv{s}$: broadly, well-fitting anchor models will be those for which the elements of $\mathscr{S}^A$ induce partitions that are supported by the data. 

A closed-form
expression for $m(\vv{y}|\vv{s})$
is available for some models, which can
provide heuristic, generalizable 
insight into which points should be anchored.  
  For example, consider a univariate location mixture model with $\sigma^2$ known,
  so that the component-specific parameter $\gamma_j$ is simply the mean of the $j$th Gaussian component.  Under the Gaussian prior
  $\pi(\vv{\gamma})= \prod_{j=1}^k \phi_1(\gamma_j; \mu, \tau^2)$ with $\mu$ and $\tau^2$ known, the
  conditional marginal likelihood satisfies the condition
  \begingroup
\vspace{-3mm} 
\small 
\begin{align} \label{conditional_marginal_likelihood}
  m(\vv{y}|\vv{s}) &\propto \exp\left(\frac{-1}{2}\sum_{j=1}^k\left(
                \sum_{i:  s_i = j}\frac{\left(  y_i -
                \bar{y}_j[\mathbf{s}] \right)^2}{\sigma^2} +\frac{
                \mu^2\tau^{-2}(1+\sigma^2)-n_j\left(\bar{y}_j[\mathbf{s}]
                -\mu\right)^2 }{(n_j\tau^2 + \sigma^2)} \right)\right)\prod_{j=1}^k
				\sqrt{\frac{n_j\tau^2 + 
                \sigma^2}{\sigma^2} },                 
                %\left(\frac{\sigma^2}{n_j\tau^2 + 
                %\sigma^2}\right) %^{1/2} 
%   \nonumber \\
%&       \hspace{27mm}  \prod_{j=1}^k\left(\frac{\sigma^2}{n_j\tau^2 + 
%                \sigma^2}\right)^{-1/2} 
\end{align}
\endgroup

\vspace{-7mm} \noindent
where $n_j = \sum_{i=1}^nI(s_i = j)$, and
$\bar{y}_j[\vv{s}] = n_j^{-1}\sum_{i:s_i=j}y_i$.

From Equation~\eqref{conditional_marginal_likelihood} we see that, for
large values of $\tau^2$ compared to $\sigma^2$, the relative magnitude of $m(\vv{y}|\vv{s})$
is determined primarily by the term $\sum_{j=1}^k \sum_{i:s_i=j}\left( y_i - \bar{y}_j[\mathbf{s}]
\right)^2$, the within-group sum of squares, for the partition induced by $\vv{s}$.  This observation
suggests a heuristic notion: well-fitting anchor models will be those
for which $\mathscr{S}^A$ contains many allocations that produce
well-separated groups in the data.  The marginal likelihood on its own
is impractical for model selection because the large cardinality of
$\mathscr{S}^A$ makes exact computation of the expression in
(\ref{anchor_marginal_likelihood}) impossible even for moderate values of
$n$ and/or $k$.  Consideration of this expression, nonetheless,
suggests that, in specifying anchor models, we should promote
separation among the mixture components.  In
Section~\ref{section::model_specification}, we propose a
computationally feasible method for specifying anchor models that
encourage separation and will \mbox{tend to fit well.} 

\vspace*{-5mm}
\subsection{Anchoring as an informative prior on
  $\vv{\gamma}, \vv{\eta}$} 
  \label{subsection::Anchor_informativeprior} 
\vspace*{-3mm}
Replacing
the exchangeable model with the anchor model
($\ref{anchor_likelihood})$ can be viewed as creating a
data-dependent, non-exchangeable prior on the component-specific
parameters.

An anchor model with anchor points $\vv{x}_1,\ldots,\vv{x}_{k_0}$
produces a posterior density of $\vv{\gamma}, \vv{\eta}$ that satisfies
\vspace{-10mm} 
\begin{align} \nonumber 
p_A(\vv{\gamma}, \vv{\eta}|\vv{y}) & \propto  \pi(\vv{\eta}) \left( \prod_{j=1}^{k_0}
                                      \pi(\vv{\gamma}_j)\phi_p(\vv{x}_j;
                                      \vv{\gamma}_j)  \right)
                                      \left( \prod_{j=k_0+1}^{k}
                                      \pi(\vv{\gamma}_j) \right) \left( \prod_{i \notin
                                      A}\sum_{j=1}^k \eta_j \phi_p(y_i;
                                      \vv{\gamma}_j)\right) \\
&=  \pi(\vv{\eta})\left(\prod_{j=1}^{k_0} C_jp(\vv{\gamma}_j |\vv{x}_j)\right)\left( \prod_{j=k_0+1}^{k}
\pi(\vv{\gamma}_j) \right) \left( \prod_{i \notin
	A}\sum_{j=1}^k \eta_j \phi_p(y_i;
\vv{\gamma}_j)\right) 
\end{align}

\vspace{-3mm} \noindent
where
$C_j = \int\pi(\vv{w})\phi_p(\vv{x}_j;
\vv{w})d\vv{w}$ and
  $p(\vv{\gamma}_j |\vv{x}_j)$ denotes the posterior density that
  results from updating the distribution of $\vv{\gamma}_j$ with the
  anchor points $\vv{x}_j$.  Because $C_j$ does not depend on $\vv{\gamma}_j$, the following proposition holds.
 
\begin{prop} 
\label{prop:proper_prior_equiv}
The anchor model described in this section
produces the same posterior distribution on $(\vv{\gamma}, \vv{\eta})$ 
as a model whose likelihood is a Gaussian mixture
on the $n-m$ unanchored observations and whose prior is equal to
$\pi(\vv{\eta})\prod_{j=1}^{k_0}p(\vv{\gamma}_j|\vv{x}_j)\prod_{j=k_0+1}^{k}\pi(\vv{\gamma}_j)$, where $p(\vv{\gamma}_j|\vv{x}_j)$ is the posterior
density of $\vv{\gamma}_j$ given the anchor points $\vv{x}_j$.
\end{prop} 
Proposition~\ref{prop:proper_prior_equiv} provides a basis for
	asymptotic results to be discussed in the next section.
\vspace*{-5mm}
\subsection{Large sample properties and quasi-consistency} 
\label{subsection:large_sample} 
\vspace*{-3mm} In this section we establish the asymptotic properties
of an anchor model and define a derived notion of {\em
  quasi-consistency\/} that enables us to quantify probabilistically
the degree of component-specific parameter identifiability attained by
an anchor model.

\vspace*{-5mm}
\subsubsection{Limiting behavior of the posterior distribution}
To characterize the limiting behavior of an anchor model, we rely on a
result from \cite{CooMac1999} which describes the large sample
properties of the posterior distribution of the model parameters
$(\vv{\gamma}, \vv{\eta})$ in a mixture model in the setting where
prior information, possibly from pre-labeled samples, is available.
Applying results of \cite{berk1966}, they derived statements that,
assuming appropriate regularity conditions, hold with probability one
with respect to the product measure $F_{\vv{\gamma}_0, \vv{\eta}_0}$
on the space of sample paths of the data-generating process with true
model parameters $(\vv{\gamma}_0, \vv{\eta}_0)$.

In addition to C.1 and C.2 on page~\pageref{item:conditions}, the
subsequent results require an additional assumption on the prior
density $\pi(\vv{\gamma})$. 
The additional assumption asks
that  %\vspace{-3mm}
\begin{enumerate} \label{item:condition_c.3}
	\item[{\bf C.3:}]  
	$\pi(\rho_q(\vv{\gamma}_0))>0, \quad \mbox{for some }q \in \{1,\ldots,k!\}$,
\end{enumerate}  that is, the prior density
is positive at some relabeling of the true parameter value. 
  Define also
$\Gamma_0 = \{ \rho_q(\vv{\gamma}_0, \vv{\eta}_0),\,\, q=1,\ldots,k!
\}$, the set containing all possible labelings of $(\vv{\gamma}_0, \vv{\eta}_0)$ .   Under C.1-C.3, the results in \cite{CooMac1999} give the following.
Let $U$ denote an arbitrary open neighborhood
of $\Gamma_0$.  Then, \vspace{-3mm}
\begin{align} \label{asymptotic_distribution1}
	\lim_{n \rightarrow \infty} \Pi(U \, | \, y_1,\ldots,y_n) = 1, \,\,\,
	a.s.-F_{\vv{\gamma}_0,\vv{\eta}_0}, 
\end{align}

\vspace{-3mm} \noindent where $\Pi(\cdot \,| \, y_1,\ldots,y_n)$ is
the posterior probability measure on the parameter space, given a random
sample $y_1,\ldots,y_n$ of size $n$.  In addition, let
$N_\epsilon(\vv{\gamma}, \vv{\eta})$ denote an open ball of radius $\epsilon > 0$
centered at $(\vv{\gamma}, \vv{\eta})$.  Consider the $q$th relabeling
$\rho_q(\vv{\gamma}_0,  \vv{\eta}_0)$ of the true model parameters, and
assume that $\epsilon$ is small enough for
$\bigcap_{h=1}^{k!} N_\epsilon(\rho_h(\vv{\gamma}_0,  \vv{\eta}_0)) = \emptyset$ to
hold.  Then,
\vspace{-3mm} 
\begin{align} \label{asymptotic_theta_distribution}
	\lim_{n \rightarrow \infty}
	\Pi(N_{\epsilon}\left(\rho_q(\vv{\gamma}_0,  \vv{\eta}_0)\right) \,| \,y_1,\ldots,y_n) =
	\frac{\pi(\rho_q(\vv{\gamma_0}, \vv{\eta}_0))}{
		\sum_{h=1}^{k!}\pi(\rho_h(\vv{\gamma}_0,\vv{\eta}_0))     }, 
	\;\;\;a.s.-F_{\vv{\gamma}_0, \vv{\eta}_0}, 
\end{align}

\vspace{-3mm} \noindent
where $\pi$ is the prior density of $(\vv{\gamma}, \vv{\eta})$. Combined, results \eqref{asymptotic_distribution1} and \eqref{asymptotic_theta_distribution} indicate that the posterior distribution concentrates on $\Gamma_0$ and, in the limit, the relative posterior mass given to the $q$th element of $\Gamma_0$ is determined solely by its prior density.  

It is natural to interpret the limiting values
	in~(\ref{asymptotic_theta_distribution}) as defining a
	discrete asymptotic probability distribution on the $k!$ elements of $\Gamma_0$, where the probabilities are determined solely by the prior.  
	Under the exchangeable model, $\pi(\cdot)$ is equal for all elements of $\Gamma_0$, and we obtain
	a discrete {\em uniform\/} distribution on its $k!$ elements: no matter
	how much additional data accumulate, all relabelings of the true
	parameter remain equally likely. 

For an anchor model, we can use the data-dependent prior given in~Proposition~\ref{prop:proper_prior_equiv} to determine the limiting values  in~\eqref{asymptotic_theta_distribution}. 
The
probability of the $q$th relabeling of $(\vv{\gamma}_0,\vv{\eta}_0)$ is in fact determined by the anchor points: it is proportional to $
\prod_{j=1}^{k_0}p(\vv{\gamma}_{0\rho_q(j)}|\vv{x_j})$. (See the proof of~\ref{pq_expression_prop}. In particular, the
expression does not depend on $\vv{\eta}_0$ because the anchor points
provide no information about the mixture proportions.) For this reason, we will use $P_{\vv{x}}(\vv{\gamma}_0) = \{ p_q,\,\, q=1,\ldots,k!  \}$ to denote the asymptotic 
distribution on $\Gamma_0$ induced by a set of anchor points $\vv{x}$, or  
$P_{\vv{x}}$ when there is no ambiguity. It is straightforward to derive expressions for the elements of 
$P_{\vv{x}}$, which depend only on the Gaussian densities of the anchor points at $\vv{\gamma}_0$, as stated in the ensuing proposition which is proved in the
Appendix.

\vspace{-3mm} 
\begin{prop} \label{pq_expression_prop}
	The $q$th element of $P_{\vv{x}}(\vv{\gamma}_0)$, $p_q$, is equal to
	\begin{align} \label{pq_expression}
	\prod_{j=1}^{k_0}\phi_p\left(\vv{x}_j; \vv{\gamma}_{0\rho_q(j)}\right)\left/
	\sum_{h=1}^{k!}\prod_{j=1}^{k_0}\phi_p\left(\vv{x}_j;\vv{\gamma}_{0\rho_h(j)}\right)\right..
	\end{align} 
\end{prop} \vspace{-3mm} 

\subsubsection{Quantifying parameter identifiability}
The result in~(\ref{asymptotic_distribution1}) states that, as the
sample size goes to infinity, the posterior mass concentrates on
arbitrarily small neighborhoods of the $k!$ relabelings of the true
value $(\vv{\gamma}_0, \vv{\eta}_0)$.  Thus, for both the exchangeable
model and the anchored model, the asymptotic distributions concentrate
on the elements of $\Gamma_0$.  The influence of the anchor points
does not disappear in large samples, however; they determine, through
$P_{\vv{x}}$, the relative posterior mass given to the $k!$ modal
regions that are treated symmetrically under the exchangeable model.
If the distribution $P_{\vv{x}}$ assigns high probability to one
relabeling, then for large samples, the density
$p_A(\vv{\gamma},\vv{\eta}|\vv{y})$ will approximate, up to a constant
factor, the density $p_E(\vv{\gamma},\vv{\eta}|\vv{y})$ around one of
the modal regions and will tend to be flat elsewhere.

The anchor points play a crucial role in disambiguating between class
labels and determining the degree with which the anchor model isolates one posterior mode.  Anchor models for which $P_{\vv{x}}$ places high probability on only one relabeling of $(\vv{\gamma}_0,\vv{\eta}_0)$ should be preferred.
 This motivates the following definition.
\begin{definition} \label{def:quasi_consistency}
  Let $P_{\vv{x}}(\vv{\gamma}_0)= \{ p_q,\,\, q=1,\ldots,k!  \}$ be the limiting probability defined above for an
  anchor model with anchor points $\vv{x}$.  We say that the anchor
  model is $\alpha$ {\em quasi-consistent\/} if
  $\max_{q \in \{1, \ldots, k!\}} p_q = \alpha$
\end{definition}

The ideal value of $\alpha$ is one; this is not attainable in practice
because all probabilities in~\eqref{pq_expression} are positive, but,
for a good anchor model, $\alpha$ will be near one and we
can report $\alpha$ as an objective measure of the quality of an
anchor model.  Note that $\alpha=\alpha(\vv{\gamma}_{0})$ depends on
the true, unknown, component-specific parameters $\vv{\gamma}_{0}$.
In practice, the latter will be replaced by an estimate
$\hat{\vv{\gamma}}_{0}$ (typically a maximum a posteriori or MAP estimate under the exchangeable model), and we would
report the value $\hat{\alpha}=\hat{\alpha}(\hat{\vv{\gamma}}_{0})$.
%Note also that, in principle, $\hat{\alpha}$ can always be calculated,
%irrespective of the method used to select the anchor points.

A related measure of how effectively the anchor points can resolve,
asymptotically, the labeling ambiguity is given by the entropy of
$P_{\vv{x}}$,
$En(P_{\vv{x}}, \vv{\gamma}_0) = -\sum_{q=1}^{k!} p_q\log(p_q)$, where
we define $\log(0) = 0$ in the expression.  Lower entropy values are
preferred. As in the
calculation of $\widehat{\alpha}$ above, we can obtain a plug-in estimate of the entropy by substituting an estimate
$\widehat{\vv{\gamma}}_{0}$ of $\vv{\gamma}_0$ into the expression.  In Section~\ref{section::Examples}, we also demonstrate a
modeling strategy in which we select anchor points to minimize the
entropy. 

The following proposition gives two interesting results in
which it can be shown analytically that certain choices of anchor
points minimize $En( P_{\vv{x}}, \vv{\gamma}_0)$ in a univariate mixture in which $\vv{\gamma}_j = (\theta_j, \sigma_j^2)$, the mean and variance of the $j$th Gaussian component.  
\vspace{-2mm}
\begin{prop} \label{prop.k2results}
	Suppose that $k=2$ and that  $m_j = m$ observations (with $1\leq m \leq n/2$) are to be anchored to component $j$, $j=1,2$.
	The following results hold: \vspace{-2mm}
	\begin{enumerate}%[noitemsep, topsep=6pt]
		\item  (Location mixture.) If $\sigma_1^2 =  \sigma_2^2= \sigma^2$ and $\theta_1 < \theta_2$, then the optimal anchoring sets
		$\vv{x}_1 = ( y_{(1)},\ldots, y_{(m)})$, $\vv{x}_2 =( y_{(n-m+1)},
		y_{(n)} )$, where $y_{(l)}$ denotes the $l$th order statistic.  %If $m_1 = m_2 = 1$, the minimum and maximum values should be chosen as anchor points. 
		\item (Scale mixture.) If $\theta_1 = \theta_2 = \theta$ and $\sigma_1^2 < \sigma_2^2$,
		then the optimal anchoring sets $\vv{x}_1$ equal to the points
		that minimize 
		$\sum_{i=1}^{m} (y_i - \theta)^2$
		and $\vv{x}_2$ equal to the
		points that maximize 
		$\sum_{i=1}^{m} (y_i - \theta)^2$. 
	\end{enumerate}
\end{prop} %\vspace{-3mm} 
\vspace{-1mm} Proposition~\ref{prop.k2results} is proved in the
Appendix and provides some intuition regarding which anchor models
most effectively alleviate the labeling ambiguity: the minimum-entropy
anchor model arises from choosing points that are as dissimilar as
possible in location and/or in scale.

\vspace*{-5mm}
 \section{Model specification} \label{section::model_specification}
\vspace*{-2mm}
We now address two fundamental issues: {\em how
    many\/} and {\em which\/} points to anchor.  
\vspace*{-5mm}
\subsection{Choosing the number of anchor
  points} \label{sec:no_of_points} \vspace*{-3mm} Strengthening
assumption C.1 by requiring equal mixture proportions (and, as a
consequence, equal probabilities on all allocation vectors), the
following proposition states that the marginal likelihood of an anchor
model can always be increased by introducing additional anchor points.
The proposition is proved in the appendix.  \vspace{-1mm}
\begin{prop} \label{prop:increasing_marg_like} Assume that $\eta_j =
  1/k$, $j=1,\ldots,k$.  
	Let $A_*^1,\ldots, A_*^n$ be a sequence of
  anchor models where $A_*^m$ has the highest marginal likelihood
  among all anchor models with $m$ anchor points. The marginal
  likelihoods of the models satisfy
  $m_{A_*^{1}}(\vv{y}) \leq \ldots \leq m_{A_*^{n}}(\vv{y})$.
\end{prop} \vspace{-1mm}  This result indicates that based on goodness-of-fit
alone, it is best to specify a larger number of optimally-anchored points.
However, increasing the number of anchor points strengthens the degree
of prior information built into the model and may increase bias in finite samples.  Intuition suggests that
limiting the number of anchor points might be desirable to ensure
satisfactory out-of-sample predictive performance. To assess the trade-off
between goodness of fit and out-of-sample predictive performance, we
conducted two small simulation studies. The details of the studies and their results are presented in the
Appendix. In simulation~1, we drew data from a
two-component location mixture and fit the optimal anchor model with varied values of
$m$.   In simulation~2, we drew data from several univariate location-scale mixtures and fit, for varied values of $m$, anchor models whose anchor points are chosen using the EM strategy to be described in Section~\ref{section:aem}.  

In agreement with our intuition, the simulation findings show that, in
cases of mixture components that are not well-separated, the
out-of-sample predictive ability of the model suffers when too many
anchor points are chosen. It is difficult to select a large
  number anchor points that accurately represent the true densities
  and they introduce bias in parameter estimation.  Further, we found
  that little benefit accrues from anchoring more than one or two
  points to each component if the components are well-separated.  
These results support the recommendation to anchor either one or
  two points to each component.
  One anchor point per component will provide unique labeling for all
  components, as given in Proposition~\ref{prop:unique_labeling}.  If
  instead each component has two anchor points, improper priors may be
  used for the component-specific features of the Gaussian mixture
  components, a specification that is impossible under the
  exchangeable model.   The next section proposes a method for
selecting which points to anchor.

\vspace*{-5mm}
\subsection{Anchored EM algorithm for selecting anchor points} \label{section:aem}
\vspace*{-3mm} We propose selecting anchor points by formulating the optimal anchor model as a solution
to a modified Bayesian expectation-maximization (EM) algorithm for
maximum a posteriori estimation.  The method proceeds by computing a
lower bound on the log posterior density of $(\vv{\gamma},\vv{\eta})$ and
iteratively updating the parameter values and an anchored posterior
distribution on the latent allocations to maximize this lower bound.
Intuitively, a good anchor model should concentrate its posterior mass
in the vicinity of one of the modal regions of the exchangeable model.
Thus, we select as the optimal anchor points those that produce the
best approximation (as measured by the lower bound) to the
exchangeable posterior density
$p_E(\vv{\gamma},\vv{\eta}|\vv{y})$
near one of the symmetric local modes.

The method draws on the formulation of \cite{Neal1998} of the
EM algorithm, which makes use of the
following lower bound on the log posterior density of $(\vv{\gamma},\vv{\eta})$:  
\vspace{-3mm} 
\begin{align}
\log\left(p_E(\vv{\gamma},\vv{\eta}|\vv{y})\right) \geq \sum_{\vv{s}\in
  \mathscr{S}}q(\vv{s})\log\left(
  \frac{p(\vv{s},\vv{\gamma}, \vv{\eta}|\vv{y})}{q(\vv{s})}\right). \label{lower
  bound} 
\end{align} 

\vspace{-3mm} \noindent
This bound holds for any distribution $q$ on the latent allocations by
Jensen's inequality.
The expression on the right-hand side of (\ref{lower bound}) is a
function of $(\vv{\gamma},\vv{\eta})$ and $q$ and will be denoted by
$F(\vv{\gamma},\vv{\eta},q)$. 
\cite{Neal1998} show that the EM algorithm may be seen as an iterative
maximization of the lower bound, $F$, with respect to $\vv{\gamma}$ (M
step) and $q$ (E step).
At iteration $t$, conditional on the current parameter value $(\vv{\gamma}^{t-1}, \vv{\eta}^{t-1})$, the distribution $q_*^t$ that maximizes
$F(\vv{\gamma}^{t-1},\vv{\eta}^{t-1},\cdot)$ is the posterior distribution on the latent
variables. For a Gaussian mixture, $q_*$ has the form
$q_*(\vv{S}=(s_1,\ldots,s_n)|\vv{y},\vv{\gamma}, \vv{\eta}) =
\prod_{i=1}^n \prod_{j=1}^k r_{ij}^{I(s_i=j)}$, 
where
\vspace{-3mm} 
\begin{align} \label{posterior allocation distribution} 
 r_{ij} =
  \frac{\eta_j\phi_p(y_i;\vv{\gamma}_j)}{\sum_{l=1}^k\eta_l\phi_p(y_i;\vv{\gamma}_l)}, \quad
  j=1,\ldots,k, \quad i=1,\ldots,n. 
\end{align}  

\vspace{-3mm} \noindent
When $q^t$ is set equal to $q_*^t = q_*(\cdot|\vv{y},\vv{\gamma}^{t-1}, \vv{\eta}^{t-1})$ in the E step of the algorithm, the
inequality in~\eqref{lower bound} is an equality for $(\vv{\gamma},\vv{\eta}) = (\vv{\gamma}^{t-1},\vv{\eta}^{t-1})$ and the lower bound is
equal to the log posterior density at that point.  Further, \cite{Neal1998} state
that the value of $(\vv{\gamma},\vv{\eta})$ that maximizes $F(\cdot, q_*^t)$ also
maximizes the log posterior density. 

\cite{Ganchev2008} have modified this EM formulation
 for settings where $q_*$ cannot arise as the
 distribution of $\vv{S}$ because the model imposes certain
restrictions on the latent variables.  
Because the lower bound on $\log( p_E(\vv{\gamma},\vv{\eta}|\vv{y}))$
 holds for any valid probability
distribution $q$, the E-step may be modified so that
$q^t$ is chosen to maximize $F(\vv{\gamma}^{t-1},\vv{\eta}^{t-1},\cdot)$, subject to
the problem-specific constraints.  It is
straightforward to verify that the lower bound on
$\log( p_E(\vv{\gamma},\vv{\eta}|\vv{y}))$ satisfies
\vspace{-8mm} 
\begin{align} 
F\left(\vv{\gamma},\vv{\eta},q\right) &= \log( p_E(\vv{\gamma},\vv{\eta}|\vv{y})) - D_{KL}(q
                              || q_* ), \label{eq:KL_bound}
\end{align} 

\vspace{-5mm} \noindent
where
$D_{KL}(q || q_*) = \sum_{\vv{s}}q(\vv{s})
\log\left(q(\vv{s})/q_*(\vv{s})\right)$ is the
Kullback-Leibler (KL) divergence of $q_*$ from $q$, and $q_*$ is the
optimal posterior distribution given in (\ref{posterior allocation
  distribution}).  For a given $(\vv{\gamma}, \vv{\eta})$, the lower bound $F$ will be largest when $q$ is as
close as possible to $q_*$, in terms of KL divergence.

An anchor model imposes constraints on the distribution of $\vv{S}$
and fits neatly into this framework. In fact,
\cite{Neal1998} and
%\cite{Ganchev2008}, and 
\cite{Lucke2016} have suggested using such EM
modifications in closely related clustering problems based on Gaussian mixtures.
For an anchor model, the 
posterior distribution of $\vv{S}$ satisfies $q^t(\vv{S}|\vv{y},\vv{\gamma}^{t-1},\vv{\eta}^{t-1})=\prod_{i=1}^nq^t(S_i=s_i|\vv{y},\vv{\gamma}^{t-1}, \vv{\eta}^{t-1})$, where $q^t$ is constrained to satisfy
\begingroup
\vspace{-5mm} 
\small
\begin{align} \label{anchored_qs}
q^t(S_i = j|\vv{y},\vv{\gamma}^{t-1}, \vv{\eta}^{t-1}) 
  &= \begin{cases} \widetilde{r}_{ij}^t, \quad & i \notin A, \\[-9pt] 
1, \quad & i \in A_j, \\[-9pt]
0, \quad & i \in A_{j'}, \quad j' \neq j. \end{cases}
\end{align}
\endgroup

\vspace{-3mm} \noindent
Here, the sets $A_j$ have
  cardinality $|A_j| = m_j$ for $j = 1 \ldots k_0$, and the
  $\widetilde{r}_{ij}^t$ are 
probabilities such that
$\sum_{j=1}^k\widetilde{r}^t_{ij}=1$ for all $i \not\in A$.  
Subject to these requirements, the form of the optimal anchor model
corresponding to a constrained distribution $q$ that minimizes the KL
divergence appearing in~\eqref{eq:KL_bound} is described in the following proposition,
which is proved in the Appendix.
\vspace{-3mm} 
\begin{prop} \label{prop:anchorq} Let $q$ be the posterior
  distribution of the allocations under an anchor model, subject to
  the restrictions in (\ref{anchored_qs}).    The KL divergence of
  $q_*$ from $q$, evaluated at a fixed value of $(\vv{\gamma}, \vv{\eta})$, is
  minimized when the sets $A_j$ are chosen to maximize
  $\sum_{j=1}^{k_0}\sum_{i \in A_j}r_{ij}$ 
  and when $\widetilde{r}_{ij} = r_{ij}$ for
    all $i \not \in A$.
\end{prop} \vspace{-3mm} 

The modified EM
algorithm will, in the E step, hold $(\vv{\gamma}^{t-1}, \vv{\eta}^{t-1})$
  constant and update $q^t$ to correspond to a valid anchor
model with the optimal anchor 
points identified in Proposition~\ref{prop:anchorq}. This amounts to including $i$ in $A_j$ if $r_{ij}$ is among the $m_j$ largest allocation probabilities to component~$j$, except in the case when $i$ satisfies this condition for some other $j^\prime \neq j$. Details on selecting the anchor points is this case are given in the Appendix.  (In our experience, this case rarely occurs in real data applications if $m$ is small relative to the sample size.)  In the
subsequent M step, $(\vv{\gamma}^t, \vv{\eta}^t)$ is updated to maximize
the lower bound, holding $q^t$ fixed at its current value. As
  in the standard EM 
  algorithm, the M step can be accomplished by maximizing $E(\log(
  p(\vv{\gamma},\vv{\eta},\vv{s},\vv{y}))$, where the expectation is taken with
  respect to $q^t$. This maximization is computationally tractable
because 
$E(\log(
  p(\vv{\gamma},\vv{\eta},\vv{s},\vv{y}))$ can be expressed as a summation of 
$k \times n$ addenda, by an argument analogous to the one used in
the proof of Proposition~\ref{prop:anchorq}.
For the models considered in the examples of
Sections~\ref{section::Examples} and~\ref{section::Falldata}
the maximizer can be derived in closed form.
The steps 
of this ``anchored EM Algorithm'' are detailed in the Appendix.

As discussed by \cite{Ganchev2008} for related approaches,  
the anchored EM algorithm 
maximizes a penalized version of the log posterior 
density, where the penalty is given by
the KL divergence of the distribution
$q_*$ corresponding
to the exchangeable model from the distribution 
$q$ corresponding to the anchored model.
Each EM iteration updates both the parameters and the distribution $q$ in
order to increase the lower bound on the log posterior density,
yielding an optimal approximation to a local mode of the exchangeable
posterior distribution.   

\subsection{Other strategies for model specification} We propose the anchored EM method as
one potential default method for automatically selecting anchor points. The anchor modeling framework admits alternative strategies based on various modeling considerations and, if available, prior information.  For example, the large-sample properties outlined in Section~\ref{subsection:large_sample} motivate choosing anchor points that minimize the entropy of the asymptotic distribution on the relabelings, a strategy which we demonstrate in Section~\ref{section::Examples}. In a recent case study, \cite{kun:per:2019} develop a method that uses diagnostic information from a base model to choose anchor points for a mixture of regressions model.
Existing classification and clustering tools can also be used to identify anchor points as representative points from the $k$ components.  Particularly applicable are a type of semi-supervised algorithms that propose strategies for introducing artificial labels to points in classification problems.  For example, the Yarowsky ``self-training'' algorithm \citep{Yarowsky:1995, Abney2004} and variants thereof use allocation probabilities to assign artificial labels iteratively until a stopping criterion is reached. 

\vspace*{-5mm}
\section{Univariate examples} \label{section::Examples}
\vspace*{-2mm}

In this section we present several examples
of the anchor modeling approach applied to univariate data sets. 
\vspace*{-4mm}
\subsection{Sampling} \label{section:sampling} In the following
examples, we specify a conditionally conjugate prior and use a Gibbs
sampler to obtain samples from the posterior distribution.  Sampling
is a well-known challenge in the exchangeable mixture model due to the
$k!$ symmetric regions of the posterior (see, e.g., \cite{celeux,
  jasra2005, geweke2007}).  A sampler which fully explores the
posterior parameter space is, in fact, one in which perfect label
switching is observed; each symmetric region must be visited with
equal frequency and ergodic averages of the component-specific
parameters should be equal.  Label switching, then, is a requisite for
convergence of the sampling algorithm, but it is difficult to ensure
this behavior, particularly in a model with well-separated components.
\cite{celeux} recommend replacing the Gibbs sampler with a simulated
tempering scheme that promotes swift movement across modal regions of
the parameter space.  Other strategies leverage the symmetry of the
posterior to improve mixing: \cite{Fruhwirth2001} proposes a sampler
that, for each sampled value of the vector of
component-specific parameters, proposes a random permutation of the
component labels that is accepted with probability one.  A
similar approach of \cite{geweke2007} augments each sampled value with
every possible relabeling of the value.

We adopted the random permutation strategy recommended by
  \cite{Fruhwirth2001}.  The resulting algorithm is a
    modification of a standard Gibbs algorithm for sampling draws from
    the posterior distribution of $(\vv{\gamma},\vv{\eta}, \vv{s})$.
    A standard Gibbs sampler would alternate between updates of
    $(\vv{\gamma},\vv{\eta})$ sampled from the conditional
    distribution of $(\vv{\gamma},\vv{\eta})$ given $(\vv{s}, \vv{y})$
    and updates of $\vv{s}$ sampled from the conditional distribution
    of $\vv{s}$ given $(\vv{\gamma},\vv{\eta},\vv{y})$.  The modified
    sampler includes an additional step after the update of
    $(\vv{\gamma},\vv{\eta})$ conditional on $(\vv{s}, \vv{y})$.
    Precisely, at iteration $t$, a value
    $(\vv{\tilde{\gamma}}^t,\vv{\tilde{\eta}}^t)$ is generated
    conditional on $(\vv{s}^{t-1}, \vv{y})$.  Then, a random
    permutation $\rho_q$ of the component labels is selected according
    to the probabilities described in the following paragraph and
    $(\vv{\gamma}^t,\vv{\eta}^t)$ is set equal to
    $\rho_q(\vv{\tilde{\gamma}}^t,\vv{\tilde{\eta}}^t)$ before
    proceeding to the update of $\vv{s}^{t-1}$.  The
    intermediate sample $(\vv{\tilde{\gamma}}^t,\vv{\tilde{\eta}}^t)$
    is discarded and only $(\vv{\gamma}^t,\vv{\eta}^t)$ is retained as
    part of the chain.

  Under the exchangeable model, each random permutation has equal
  probability of being sampled.  Under the anchor model, the asymmetry
  of the posterior density requires the permutations to be accepted
  with probabilities proportional to the values of the
  posterior density
$p_A(\rho_q(\vv{\tilde{\gamma}}^t,\vv{\tilde{\eta}}^t) | \vv{y})$
  at the various
  relabelings. Simple algebraic manipulations show that, at
iteration $t$, permutation
  $q$ is selected with probability $w_{tq}$, for $q=1,\ldots,k!$,
  where $w_{tq}$ is equal to
\begin{align} 
w_{tq} &= \frac{\prod_{j=1}^k\prod_{i\in
         A_j}\phi(y_i;\tilde{\vv{\gamma}}_{\rho_q(j)}^t)}{\sum_{h=1}^{k!}
         \prod_{j=1}^k\prod_{i\in  
         A_j}\phi(y_i;\tilde{\vv{\gamma}}_{\rho_h(j)}^t)}, \quad q=1,\ldots,k!. 
\end{align}
We prove in the appendix that the accepted permuted draw
  $(\vv{\gamma}^t,\vv{\eta}^t)$ is in fact a draw from the
  distribution of $(\vv{\gamma},\vv{\eta})$ conditional on
  $(\vv{s}^{t-1}, \vv{y})$.  Therefore this procedure results in an
  algorithm whose invariant distribution is the same as that
  of the standard algorithm.

In each of the following examples, we used a Gibbs sampler with the
random permutation step.  We ran 50 chains initialized at different
random starting points, discarding the first 1,000 iterations of each
chain as burn-in and thinning the chains every 100-th draw to
obtain a total of 15,000 retained posterior samples.  We estimated
standard errors of the computed
  estimates using the overlapping batch means estimator implemented in
the R package \texttt{mcmcse} \citep{R:mcmcse}.

\subsection{Galaxies}
\label{section:galaxies}
Our first example demonstrates the anchoring method using the
 galaxies data set from \cite{Roeder1990},
  now a benchmarking staple of the mixture literature.  Each of the 82 measurements gives the velocity of a galaxy sampled from the Corona Borealis region of space.  Previous
analyses have indicated that between three and seven Gaussian
components are appropriate for these data; we chose to set $k=6$, the value given the highest posterior probability in the well-known analysis of \cite{RicGre1997} and used in subsequent analyses by \cite{stephens2000} and \cite{rodriguez}. 
We specify the following priors and hyperparameters following the recommendations of \cite{RicGre1997}: the parameters
$\vv{\gamma}_j = (\theta_j,\sigma_j^{2})$ are a priori mutually independent.  The $\theta_j$ are normal with mean $\mu$ and variance
$1/\kappa$, the precisions, $\sigma_j^{-2}$, have a Gamma$(a_0,b_0)$
distribution, where the Gamma distribution is parameterized to
have a mean of $b_0/a_0$, and $\vv{\eta}$
has a Dirichlet$(\vv{1}_k)$ distribution, where $\vv{1}_k$ is a vector of $k$ ones.  The hyperparameters were set as $\mu = 21.7255$, the midpoint of the data, $\sqrt{\kappa} = 1/52$, and $a_0=2$. We specified a Gamma$(0.2,0.016)$ distribution for $b_0$.

We used two methods to specify anchor models with one anchor point per component. The first model used anchor points selected by the anchored EM method described in Section~\ref{section:aem}.  We ran the anchored EM algorithm from 50 random starting positions and selected the anchor points that corresponded to the highest value of the lower bound in~\eqref{eq:KL_bound}.  

The second set of anchor points were selected to minimize the estimated entropy of the asymptotic probability distribution on the class relabelings as given by~\eqref{pq_expression}.
To do this, we treated $En(P_{\vv{x}}, \vv{\widehat{\gamma}}_0)$ as
a continuous function of $\vv{x}$, where $\vv{\widehat{\gamma}}_0$ is a maximum a posteriori estimate of 
$\vv{\gamma}_0$ under the exchangeable model, and found a value $\vv{x}^*$ 
that minimizes this continuous function. The value $\vv{x}^*$ was identified using the \texttt{optim} function in
R with the BFGS method and the tolerance parameter set equal to
$10^{-10}$.  We then chose the anchor points to be
the observations closest to $\vv{x}^*$.

We used a Gibbs sampler to fit
the two anchor models and the exchangeable model using the random permutation strategy described in Section~\ref{section:sampling}.  We applied two
popular relabeling algorithms to the samples from the exchangeable model: the KL
method \citep{stephens2000} and data-based (DB) relabeling
\citep{rodriguez}. Both methods are implemented in the R
\texttt{label.switching} package \citep{R_labelswitching}.  Lastly, we applied an ordering constraint that required $\theta_1<\ldots <\theta_6$ with probability~1. % right panel of Figure~\ref{fig:galaxies} shows the time in seconds used for the pre-processing of the two anchor models and the post-processing of the two relabeling methods.  

\begin{figure}[t!]
	\includegraphics[width=.97\textwidth]{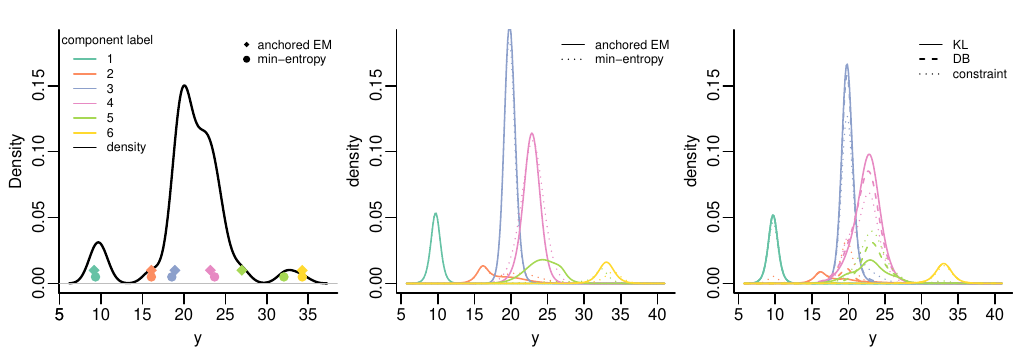}
	\caption{\label{fig:galaxies}
		Left: kernel density estimate of the galaxies data with the EM and minimum-entropy anchor points.  Middle: estimated scaled component densities under the two anchor models.  Right: estimated scaled component densities under the two relabeling methods.}
	\end{figure}

The left panel of Figure~\ref{fig:galaxies}
shows the locations of the minimum-entropy anchor points and the anchored EM points.   The EM anchor points are close to the minimum-entropy anchor points for most components except for component~5, where the minimum-entropy point falls in a cluster of high velocity near that of component~6.  Using the maximum a posteriori estimate of $\vv{\gamma}_0$, we calculated the estimated coefficients of quasi-consistency of
Definition~\ref{def:quasi_consistency}, $\widehat{\alpha}$, to be greater than 0.9999 for both of the anchor models, indicating a high degree of asymptotic identifiability of the component labels. 

The middle and right panels of Figure~\ref{fig:galaxies} display Monte Carlo estimates of the scaled component densities. The scaled component density for $j$ is $\eta_j\phi_p(\cdot,\gamma_j)$; as noted by \cite{rodriguez}, accurate posterior inference on this quantity is tantamount to inference on the posterior classification probability to component~$j$.  For components~1, 3, 4, and~6, all methods produce unimodal densities that are similar in location  The estimated density of component~2 has an irregular shape for all methods: the density is skewed under the anchored models and KL relabeling and it is bimodal under the ordering constraint.  The location of component~5 differs between the two anchor models, with the minimum-entropy model placing the component in the right-hand cluster in the data.  The  relabeling methods estimate a more symmetric density with a mode shifted towards the right.  

The posterior means of the component-specific parameters are given in Table~\ref{table:galaxies}. The anchored models and relabeling methods produce similar estimates for $\theta_1$, $\theta_3$, and $\theta_4$.  For component~4, however, the estimated standard deviations somewhat between the anchor models and relabeling methods, with the estimated values of $\sigma_4$ higher under the relabeling methods than the EM anchor model.
Both anchor models estimate a greater degree of separation among the means of components~2, 3, 4, and~5  than either of the relabeling methods, which reflects the influence of the separation among the anchor points.  Lastly, the parameter estimates for components~5 and~6 under the minimum-entropy model reflect the proximity of the anchor points assigned to these components: both components describe the distinct modal region between 30 and 35, whereas component~5 overlaps with components in the middle region under the other methods.   The estimated $\vv{\eta}$ values are fairly consistent across methods, except for a very low estimate of $\eta_5$ under the minimum-entropy model and a tendency for the ordering constraint to produce estimates closer to $1/k$ than the other methods.

% latex table generated in R 4.0.0 by xtable 1.8-4 package
% Tue Jun 30 01:26:21 2020
\begin{table}[ht]
	\caption{\label{table:galaxies}  Posterior means (standard errors) of the component-specific parameters for the galaxies data.}
	\centering
	\small
\resizebox{\textwidth}{!}{	\begin{tabular}{rlllllll}
		\hline
		& $\theta_1$ &  $\theta_2$ &  $\theta_3$ &  $\theta_4$ &  $\theta_5$ &  $\theta_6$ & time(s) \\ 
		\hline
anchored EM & 9.713 (0.0022) & 16.798 (0.0129) & 19.845 (0.0018) & 22.803 (0.0038) & 25.408 (0.0130) & 33.018 (0.0058) & 11.91 \\ 
min-entropy & 9.713 (0.0022) & 17.289 (0.0214) & 19.815 (0.0021) & 22.927 (0.0036) & 31.512 (0.0181) & 33.672 (0.0148) & 58.62 \\ 
KL & 9.711 (0.0024) & 18.522 (0.0887) & 19.847 (0.0169) & 22.625 (0.0057) & 23.419 (0.0802) & 32.795 (0.0223) & 183.46 \\ 
DB & 9.711 (0.0024) & 18.177 (0.0853) & 20.049 (0.0205) & 22.222 (0.0130) & 23.963 (0.0817) & 32.797 (0.0195) & 9.04 \\ 
constraint & 8.099 (0.0529) & 16.437 (0.0264) & 19.878 (0.0118) & 22.248 (0.0119) & 25.631 (0.0288) & 34.625 (0.0553) & N/A \\ 
		\hline
		\hline
		& $\sigma_1$ &  $\sigma_2$ &  $\sigma_3$ &  $\sigma_4$ &  $\sigma_5$ &  $\sigma_6$&\\
		\hline
anchored EM & 0.685 (0.0018) & 1.104 (0.0058) & 0.756 (0.0013) & 1.110 (0.0024) & 1.289 (0.0050) & 1.097 (0.0038) &  \\ 
min-entropy & 0.682 (0.0018) & 1.253 (0.0083) & 0.755 (0.0017) & 1.490 (0.0031) & 1.214 (0.0076) & 1.134 (0.0069) &  \\ 
KL & 0.708 (0.0037) & 1.131 (0.0048) & 0.794 (0.0031) & 1.610 (0.0020) & 1.315 (0.0045) & 1.171 (0.0031) &  \\ 
DB & 0.709 (0.0020) & 1.080 (0.0055) & 0.825 (0.0030) & 1.741 (0.0065) & 1.197 (0.0048) & 1.178 (0.0051) &  \\ 
constraint & 0.749 (0.0031) & 0.987 (0.0047) & 1.078 (0.0058) & 1.369 (0.0056) & 1.359 (0.0048) & 1.185 (0.0056) &  \\ 
		\hline
		\hline
		& $\eta_1$ &  $\eta_2$ &  $\eta_3$ &  $\eta_4$ &  $\eta_5$ &  $\eta_6$ &\\ 
  \hline
anchored EM & 0.090 (0.0002) & 0.055 (0.0005) & 0.374 (0.0007) & 0.330 (0.0009) & 0.105 (0.0008) & 0.046 (0.0002) &  \\ 
min-entropy & 0.091 (0.0002) & 0.078 (0.0011) & 0.348 (0.0007) & 0.415 (0.0011) & 0.038 (0.0003) & 0.029 (0.0002) &  \\ 
KL & 0.090 (0.0002) & 0.041 (0.0003) & 0.323 (0.0009) & 0.404 (0.0011) & 0.097 (0.0007) & 0.045 (0.0002) &  \\ 
DB & 0.090 (0.0002) & 0.052 (0.0005) & 0.312 (0.0009) & 0.376 (0.0013) & 0.124 (0.0010) & 0.046 (0.0002) &  \\ 
constraint & 0.082 (0.0003) & 0.103 (0.0010) & 0.294 (0.0013) & 0.301 (0.0013) & 0.178 (0.0014) & 0.042 (0.0002) &  \\ 
\hline
	\end{tabular} }
\end{table}

Table~\ref{table:galaxies} also reports the CPU times for implementing each method using an Intel Core i7-8700 processor.  For the anchor models, these times are the pre-processing times required to select anchor points.  For the relabeling methods, the reported times are for post-processing of the posterior samples and are thus dependent on the number of posterior samples retained for the analysis.  None of the reported times includes the time required to sample from the posterior distributions.

\subsection{Simulated data} \label{section:simexamples}
We next fit anchor models to data generated from three univariate Gaussian mixtures with the parameters given in Table~\ref{sim_models}.

\begin{table}[t!]
	\caption{\label{sim_models} Model parameters used in the simulations.}
\begin{tabular}{p{3mm}p{8mm}p{25mm}p{4cm}p{4cm}}
	&& Model 1 & Model 2  &  Model 3\\
	&$\vv{\eta}$  & $(0.65, 0.35)$ & $(0.25, 0.25, 0.25, 0.25)$ & $(0.2,0.2,0.25,0.2,0.15)$ \\
	&$\vv{\theta}$ &$(0,0)$ & $(-3,-1,1,3)$ & $(19, 19, 23, 29, 33)$ \\
	&$\vv{\sigma}$ & $(0.5, 1.5)$ & $(1,1,1,1)$& $(2.236, 1, 1, 0.707, 1.414)$ \\
\end{tabular} 
\end{table}
\vskip 3mm

\noindent The density functions of models~1, 2, and~3 are shown in the left panels of Figures~\ref{fig:model1},~\ref{fig:model2}, and \ref{fig:model3}, respectively.
Model~1 is a scale mixture whose two components have identical locations.  Models~2 and~3 have been studied previously by~\cite{ECR} (Model 3) and \cite{rodriguez} (Models~2 and~3) to assess performance of relabeling algorithms.   
We drew samples of size $n=200$ from Models~1 and~2 and $n=600$ from Model~3.  Following the approach of \cite{rodriguez}, we used ``perfect samples,'' evenly-spaced quantiles of the mixture distribution as evaluated using the R package nor1mix \citep{R:nor1mix}, to eliminate sampling variability.  

 For each model, we fit three anchor models using the anchored EM method and the  minimum-entropy method described in Section~\ref{section:galaxies}.  The third anchor model is an ``oracle'' model, in which the anchor points are selected to be the observations closest to predetermined quantiles of each true component density.  Such information about the true densities would not be available in practice, but we present these results as an illustration of the model's performance when anchor points represent known features of the true mixture components.  For the anchor models, we report the value of $\alpha$ evaluated at the true $\vv{\gamma}_0$.  Finally, we fit the exchangeable model and applied the KL and DB relabeling methods and prior ordering constraints.
 
 The following priors and hyperparameters were specified for each example, adhering to the recommendations of \cite{RicGre1997}: $\theta_j$ has a Normal distribution with mean $\mu$ and variance
 $1/\kappa$, $\sigma_j^{-2}$ has a Gamma$(a_0,b_0)$
 distribution, $j=1,\ldots,k$, and $\vv{\eta}$
 has a Dirichlet$(\vv{1}_k)$ distribution.  Letting $R=y_{(n)}-y_{(1)}$, where $y_{(h)}$ denotes the $h$th order statistic, the hyperparameters were set as: $\mu = \bar{y}$, $\kappa=1/R^2$, $a_0=2$, $b_0 \sim$Gamma($g_0$, $h_0$), $g_0=0.2$, and $h_0=10/R^2$. 
 
  The following results use $m_j=1$, $j=1,\ldots,k$ (one anchor point per component) for the three anchor models.  The oracle anchor points are the observations closest to the median of each component.   The Appendix presents results from using two anchor points per component.

\subsubsection{Model 1.}
  \begin{figure}[t!]
 	\includegraphics{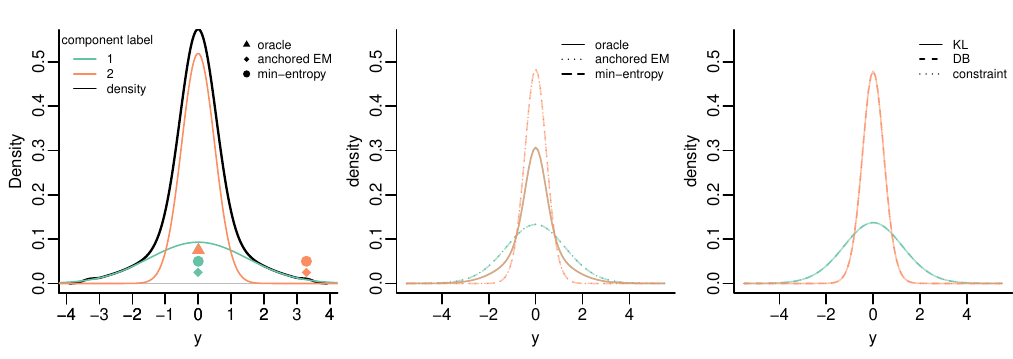}
 	\caption{\label{fig:model1} Left panel: mixture density, anchor points, and true scaled component densities for model~1. Middle panel: estimated scaled component densities for the anchor models.  Right panel: estimated scaled component densities for the relabeling methods.}
 \end{figure}
 Model~1 is a scale mixture whose two components both have means of
 zero.  The oracle anchors are nearly identical points at the median
 of the distribution.  The anchored EM and minimum-entropy methods
 select the same anchor points in this example: the maximum
 observation is anchored to component~1 while the observation closest
 to the sample mean is anchored to component~2.  These two anchor
   models have $\alpha$ values of
      1.000, while the oracle anchor model has an
  $\alpha$ value of 
   0.500.  Figure~\ref{fig:model1} shows the anchor points and the
 estimated scaled component densities, and Table~\ref{table:model1}
 gives the estimated posterior means of the component-specific
 parameters and the total absolute errors, calculated as
 $\sum_{j=1}^k|a_j-\widehat{a}_j|$ for a parameter $a_j$ having
 posterior mean equal to $\widehat{a}_j$. For this example, the
 ordering constraint $\sigma_1>\sigma_2$ was used.  The estimates are
 very similar across the methods for these data, with the exception of
 the oracle anchor model.  The oracle model performs poorly, producing
 identical parameter estimates for both components, because the anchor
 points provide no information about the scale difference between the
 components. None of the methods accurately capture the difference
 between $\eta_1$ from $\eta_2$, with the anchored EM and
 minimum-entropy models coming the closest of the six methods.

% latex table generated in R 4.0.0 by xtable 1.8-4 package
% Mon Jun 29 09:46:22 2020
\begin{table}[t!] \caption{\label{table:model1} Posterior means (standard errors) of the component-specific parameters for Model~1.}
	\centering
	\small
	\begin{tabular}{rllll}
		\hline
		& $\theta_1$ & $\theta_2$ & error & time (s) \\  
		\hline
		true & 1.5 & 0.5 & N/A &  \\ 
  oracle anchors & 0.0010 (0.0010) & -0.0010 (0.0010) & 0.0020 & N/A \\ 
anchored EM & -0.0027 (0.0014) & -0.0009 (0.0005) & 0.0036 & 1.03 \\ 
min-entropy & -0.0003 (0.0014) & -0.0006 (0.0005) & 0.0009 & 0.50 \\ 
KL & 0.0007 (0.0013) & 0.0005 (0.0005) & 0.0012 & 584.25 \\ 
DB & 0.0010 (0.0013) & 0.0002 (0.0006) & 0.0012 & 574.00 \\ 
constraint & 0.0011 (0.0013) & 0.0001 (0.0006) & 0.0012 & N/A \\ 
		\hline
		\hline
		& $\sigma_1$ & $\sigma_2$ &  error& \\ 
		\hline
		true & 0.35 & 0.65 & N/A &  \\ 
  oracle anchors & 0.867 (0.0035) & 0.866 (0.0035) & 0.9988 &  \\ 
anchored EM & 1.320 (0.0013) & 0.467 (0.0006) & 0.2125 &  \\ 
min-entropy & 1.320 (0.0013) & 0.468 (0.0006) & 0.2122 &  \\ 
KL & 1.309 (0.0013) & 0.464 (0.0006) & 0.2272 &  \\ 
DB & 1.309 (0.0013) & 0.464 (0.0006) & 0.2272 &  \\ 
constraint & 1.309 (0.0013) & 0.464 (0.0006) & 0.2272 &  \\ 
		\hline
		\hline
		& $\eta_1$ & $\eta_2$ & error & \\  
		\hline
		true & 0.35 & 0.65 & N/A &  \\ 
  oracle anchors & 0.501 (0.0009) & 0.499 (0.0009) & 0.302 &  \\ 
anchored EM & 0.430 (0.0009) & 0.570 (0.0009) & 0.1594 &  \\ 
min-entropy & 0.428 (0.0009) & 0.572 (0.0009) & 0.1568 &  \\ 
KL & 0.438 (0.0009) & 0.562 (0.0009) & 0.1766 &  \\ 
DB & 0.439 (0.0009) & 0.561 (0.0009) & 0.1772 &  \\ 
constraint & 0.439 (0.0009) & 0.561 (0.0009) & 0.1771 &  \\ 
		\hline
	\end{tabular}
\end{table}

\subsubsection{Model 2.} 
 \begin{figure}[ht]
	\includegraphics{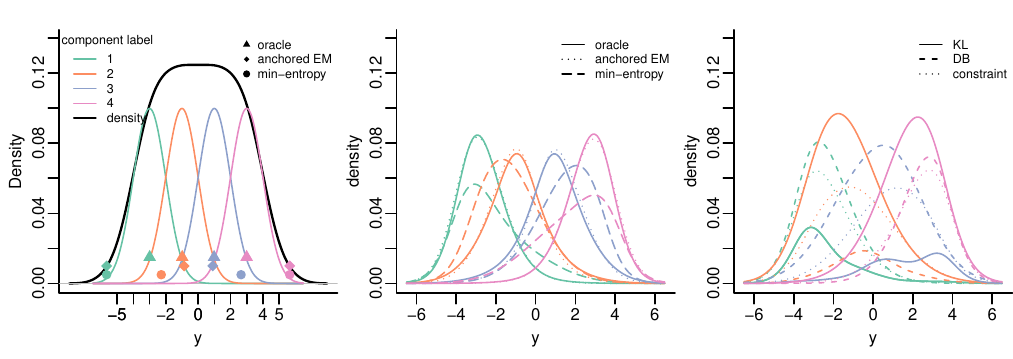}
	\caption{\label{fig:model2} Left panel: mixture density, anchor points, and true scaled component densities for model~2. Middle panel: estimated scaled component densities for the anchor models.  Right panel: estimated scaled component densities for the relabeling methods.}
\end{figure} 
The component densities of Model~2 overlap substantially, with equal
variances and evenly-spaced means. The true model and estimated anchor
points are shown in the left panel of Figure~\ref{fig:model2},
with $\alpha$ values of 0.947,
  0.972, and 0.996 for the oracle, anchored EM, and minimum-entropy
  models, respectively.  The EM and minimum-entropy method both
select the minimum and maximum observations to be anchored to
components~1 and~4, respectively.  For components~2 and~3, the
anchored EM points fall near the true component means of -1 and 1,
while the minimum-entropy points fall closer to -2 and 2,
respectively. The effect of these differences is seen in the middle
panel of Figure~\ref{fig:model2}. The estimated scaled densities for
components~2 and~3 are approximately symmetric under the EM anchor
model.  Under the minimum entropy model, however, these densities are
skewed, with excess mass near the anchor points, and the posterior
means of $\theta_2$ and $\theta_3$, shown in Table~\ref{table:model2},
are poor estimates of the true component means.  Both of the
relabeling methods produce estimated component means that are biased
towards zero, most substantially for components~2 and~3, and severely
overestimate $\eta_2$ while underestimating $\eta_3$.  The oracle and
the EM anchor models produce relatively accurate estimates of all
model parameters.

% latex table generated in R 4.0.0 by xtable 1.8-4 package
% Wed Jun  3 10:36:59 2020
\begin{table}[t!]
	\caption{\label{table:model2} Posterior means (standard errors) of the component-specific parameters for Model~2. }
	\small
	\centering
\resizebox{\textwidth}{!}{	\begin{tabular}{rllllll}
		\hline
		& $\theta_1$ & $\theta_2$ & $\theta_3$ & $\theta_4$ & error & time (s) \\ 
		\hline
		\hline
		true & -3 & -1 & 1 & 3 &  & \\ 
  oracle anchors & -2.819 (0.0049) & -0.960 (0.0072) & 0.956 (0.0072) & 2.818 (0.0049) & 0.448 & N/A \\ 
anchored EM & -2.983 (0.0069) & -0.960 (0.0073) & 0.997 (0.0073) & 2.997 (0.0068) & 0.063 & 24.87 \\ 
min-entropy & -2.967 (0.0116) & -1.653 (0.0095) & 1.965 (0.0088) & 2.864 (0.0134) & 1.786 & 10.37 \\ 
KL & -1.773 (0.0344) & -1.529 (0.0084) & 1.194 (0.0339) & 2.074 (0.0069) & 2.876 & 7031.77 \\ 
DB & -2.556 (0.0142) & -0.250 (0.0384) & 0.260 (0.0143) & 2.513 (0.0196) & 2.421 & 169.71 \\ 
constraint & -3.483 (0.0231) & -1.038 (0.0121) & 1.034 (0.0124) & 3.453 (0.0231) & 1.008 & N/A \\ 
		\hline
		\hline
		& $\sigma_1$ & $\sigma_2$ & $\sigma_3$ & $\sigma_4$ & error & \\ 
		\hline
		true &    1 &    1 &    1 &    1 & & \\ 
  oracle anchors & 1.024 (0.0026) & 1.067 (0.0041) & 1.063 (0.0039) & 1.026 (0.0025) & 0.1805 &  \\ 
anchored EM & 1.074 (0.0025) & 0.984 (0.0031) & 0.984 (0.0032) & 1.069 (0.0025) & 0.1752 &  \\ 
min-entropy & 1.121 (0.0040) & 1.089 (0.0036) & 1.042 (0.0034) & 1.168 (0.0043) & 0.4192 &  \\ 
KL & 0.749 (0.0067) & 1.449 (0.0023) & 0.715 (0.0069) & 1.304 (0.0031) & 1.2890 &  \\ 
DB & 1.077 (0.0033) & 0.698 (0.0028) & 1.455 (0.0051) & 0.987 (0.0032) & 0.8471 &  \\ 
constraint & 0.984 (0.0038) & 1.124 (0.0047) & 1.122 (0.0047) & 0.988 (0.0038) & 0.2733 &  \\ 
		\hline
		\hline
		& $\eta_1$ & $\eta_2$ & $\eta_3$ & $\eta_4$  & error & \\ 
		\hline
		true & 0.25 & 0.25 & 0.25 & 0.25 & & \\ 
  oracle anchors & 0.247 (0.0011) & 0.252 (0.0013) & 0.251 (0.0013) & 0.250 (0.0010) & 0.0059 &  \\ 
anchored EM & 0.255 (0.0012) & 0.247 (0.0012) & 0.247 (0.0012) & 0.251 (0.0012) & 0.0124 &  \\ 
min-entropy & 0.222 (0.0016) & 0.283 (0.0016) & 0.263 (0.0014) & 0.232 (0.0018) & 0.0923 &  \\ 
KL & 0.100 (0.0007) & 0.439 (0.0016) & 0.082 (0.0006) & 0.379 (0.0015) & 0.6357 &  \\ 
DB & 0.275 (0.0015) & 0.078 (0.0007) & 0.418 (0.0022) & 0.230 (0.0014) & 0.3844 &  \\ 
constraint & 0.224 (0.0017) & 0.275 (0.0020) & 0.274 (0.0019) & 0.226 (0.0017) & 0.0992 &  \\  
		\hline
	\end{tabular} }
\end{table}

\subsubsection{Model 3} 
\begin{figure}[ht]
	\includegraphics{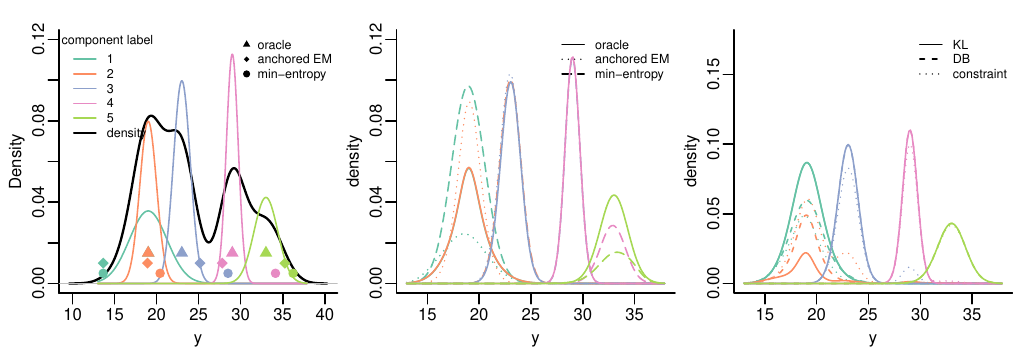}\caption{\label{fig:model3} Left panel: mixture density, anchor points, and true scaled component densities for model~3. Middle panel: estimated scaled component densities for the anchor models.  Right panel: estimated scaled component densities for the relabeling methods. }
\end{figure}

Model~3 is a five-component mixture in which the components have
varying locations, scales, and weights, except components~1 and~2,
which have identical locations.  The true scaled component densities
and anchor points are shown in Figure~\ref{fig:model3} and the
posterior parameter estimates are given in Table~\ref{table:model 3}.
The$\alpha$ values for the
  oracle, anchored EM, and minimum-entropy models are 0.500, 1.000,
  and 0.978, respectively.  The anchored EM points are offset from
the true component means but still tend to fall in regions to which
their component's true density assigns sizable mass.  This model
produces accurate estimates of $\theta_2 - \theta_5$, but incorrectly
estimates a substantial location shift between components~1 and~2.
Its estimates of the component standard deviations are close to their
true values, except that of $\sigma_1$, which is underestimated as it
is by all other methods.  The KL and DB relabeling methods accurately
estimate the means and standard deviations of all components, and the
DB method also produces accurate estimates of $\vv{\eta}$.  Under the
ordering constraint, the estimated scaled densities of components~2
and~3 are multimodal and the estimates of $\vv{\theta}$ are
comparatively inaccurate, reflecting the inadequacy of this constraint
to describe overlapping components.
% latex table generated in R 4.0.0 by xtable 1.8-4 package
% Wed Jun  3 16:08:52 2020
\begin{table}[ht]
	\small
	\caption{\label{table:model 3} Posterior means (standard errors) of the component-specific parameters for Model~3.}
	\centering
\resizebox{\textwidth}{!}{	\begin{tabular}{rlllllll}
		\hline
		& $\theta_1$ &$\theta_2$ & $\theta_3$ & $\theta_4$ & $\theta_5$ & error & time (s) \\ 
  \hline
true & 19 & 19 & 23 & 29 & 33 &  & \\ 
  oracle anchors & 18.955 (0.0049) & 18.958 (0.0049) & 23.030 (0.0014) & 29.006 (0.0006) & 32.990 (0.0016) & 0.135 & N/A \\ 
anchored EM & 17.273 (0.0290) & 19.049 (0.0019) & 22.975 (0.0014) & 29.006 (0.0006) & 32.991 (0.0016) & 1.816 & 16.60 \\ 
min-entropy & 18.948 (0.0014) & 23.011 (0.0010) & 29.006 (0.0006) & 33.031 (0.0056) & 33.838 (0.0100) & 14.94 & 10.56 \\ 
KL & 19.080 (0.0028) & 20.037 (0.0876) & 23.029 (0.0016) & 29.006 (0.0007) & 32.993 (0.0018) & 1.159 & 1106.45 \\ 
DB & 20.376 (0.0698) & 18.732 (0.0417) & 23.035 (0.0018) & 29.007 (0.0007) & 32.995 (0.0018) & 1.691 & 287.54 \\ 
constraint & 17.808 (0.0288) & 20.061 (0.0299) & 23.847 (0.0360) & 29.262 (0.0122) & 33.167 (0.0146) & 3.529 & N/A \\
\hline
%	\end{tabular}
%\begin{tabular}{rllllll}
	\hline
	& $\sigma_1$ & $\sigma_2$ & $\sigma_3$ & $\sigma_4$ & $\sigma_5$ & error & \\ 
  \hline
true & 2.236 &    1 &    1 & 0.7071 & 1.414 & & \\ 
  oracle anchors & 1.371 (0.0048) & 1.380 (0.0048) & 0.973 (0.0010) & 0.719 (0.0005) & 1.364 (0.0012) & 1.334 &  \\ 
anchored EM & 1.490 (0.0081) & 1.073 (0.0042) & 1.011 (0.0009) & 0.719 (0.0005) & 1.363 (0.0012) & 0.892 &  \\ 
min-entropy & 1.606 (0.0010) & 1.011 (0.0007) & 0.718 (0.0005) & 1.067 (0.0025) & 1.134 (0.0037) & 1.563 &  \\ 
KL & 1.582 (0.0047) & 0.976 (0.0051) & 0.974 (0.0011) & 0.716 (0.0005) & 1.358 (0.0013) & 0.769 &  \\ 
DB & 1.538 (0.0080) & 1.020 (0.0053) & 0.974 (0.0011) & 0.716 (0.0005) & 1.358 (0.0013) & 0.810 &  \\ 
constraint & 1.325 (0.0050) & 1.265 (0.0061) & 0.934 (0.0022) & 0.745 (0.0018) & 1.337 (0.0019) & 1.356 &  \\  
\hline
%\end{tabular}
%\begin{tabular}{rllllll}
	\hline
	& $\eta_1$ & $\eta_2$ & $\eta_3$ & $\eta_4$ & $\eta_5$ & error & \\ 
  \hline
true & 0.2 & 0.2 & 0.25 & 0.2 & 0.15 & & \\ 
  oracle anchors & 0.201 (0.0008) & 0.204 (0.0008) & 0.245 (0.0003) & 0.201 (0.0001) & 0.149 (0.0001) & 0.0127 &  \\ 
anchored EM & 0.132 (0.0021) & 0.255 (0.0019) & 0.264 (0.0003) & 0.201 (0.0001) & 0.149 (0.0001) & 0.1387 &  \\ 
min-entropy & 0.392 (0.0002) & 0.255 (0.0002) & 0.200 (0.0002) & 0.088 (0.0004) & 0.064 (0.0004) & 0.4954 &  \\ 
KL & 0.335 (0.0010) & 0.073 (0.0008) & 0.246 (0.0005) & 0.198 (0.0002) & 0.147 (0.0002) & 0.2691 &  \\ 
DB & 0.257 (0.0015) & 0.151 (0.0018) & 0.246 (0.0005) & 0.198 (0.0002) & 0.147 (0.0002) & 0.1146 &  \\ 
constraint & 0.191 (0.0023) & 0.252 (0.0012) & 0.228 (0.0010) & 0.186 (0.0007) & 0.143 (0.0003) & 0.1034 &  \\  
\hline
\end{tabular} }
\end{table}

The minimum entropy model performs poorly for these data: the anchor points are located at the periphery of plausible regions under their respective component densities, with no points near regions of high density around the mean of component~3.  As a result, the mass at this peak is split between components~2 and~3 and the estimated scaled component densities for components~2 and~3 appear bimodal.  A similar phenomenon produces bimodality in component~4.  Table~\ref{table:model 3} shows that this model produces the least accurate estimates of all of the methods, especially for the component means.  

In each of these univariate examples, the anchored EM algorithm has selected anchor points that result in comparatively accurate estimates of the component-specific parameters.  In terms of absolute relative error, it outperforms the minimum-entropy anchor model in all cases and tends to have comparable or superior performance to the relabeling methods.  Interestingly, the anchored EM model occasionally performs better than the oracle model, due to the anchor points' ability  to provide information about both the locations and the relative scales of the components.  The estimates produced by the minimum entropy model are less accurate because this method tends to select points in low-density areas where adjacent component begin to overlap.   Although these points maximize the model's estimated asymptotic identifiability, their influence introduces  bias in finite samples. It is evident that anchor points must fall in areas with non-negligible density of the components to which they belong.  

%The relabeling methods also produce accurate estimates for most cases, although the posterior means have much more variability.  This may be due to the posterior uncertainty produced by the possibility of empty components under the exchangeable model. \todo{think about this and whether it is true.} 

%\clearpage
\vspace*{-8mm}
\section{A multivariate example: fall detection data}
  \label{section::Falldata} 
  \vspace*{-2mm} We now apply the anchored modeling framework to a
  data set called SisFall \citep{Sucetal2017}, one of a growing body
  of fall data sets that are being used to develop systems that detect
  falls automatically using wearable devices, cameras, and/or
  microphones.  Experimental data are obtained from volunteer subjects
  who simulate falls and various activities of daily living (ADLs) and
  analyzed with the goal of characterizing the distinguishing features
  of falls compared to ADLs and detecting falls with
  high accuracy.
% The subjects of the SisFall experiments performed $15$ types of
% falls and $15$ types of ADLs while
% wearing two accelerometers and one
% gyroscope.
Common practices in analyzing these types of data include
thresholding \citep{bourke2008}, in which lower- or upper-thresholds
for one variable are set, and a fall is determined to have occurred if
the variable exceeds the threshold during a trial. More recent
analyses have used supervised classification algorithms on extracted
features of the data \citep{Marketal2012, Casetal2017}.  Our approach
uses a finite Gaussian mixture model to cluster activities into
similar subgroups and to provide a characterization of the features of
each group. Analyzing these data in a mixture framework makes it
possible to identify groups of experimental activities that share
similar features and to describe, with an accompanying
  appraisal of
uncertainty, the 
typical features of each group.  Using this model for classification
can provide further insight about which types of ADLs are difficult to
distinguish from falls.

The subjects of the SisFall experiments performed $15$ types of falls
and $19$ types of ADLs, repeating $5$ trials for most of the activities, while wearing two
accelerometers and one gyroscope.  We analyzed the data
recorded by one of the two accelerometers worn by
 one subject (``Subject 9'', a
  24-year-old male) in the SisFall dataset.  A time series of
three-dimensional acceleration vectors $(x_t, y_t, z_t)$ is available
for each of the $154$ trials.  
Following common practice in the fall detection literature, we
summarized the acceleration at each time point $t$ via the
Signal Magnitude Vector (SMV), defined as
$SMV_t = \sqrt{x_t^2 + y_t^2 + z_t^2}$.  We further summarized the
SMV series for each trial using  the logarithm of three extracted features
arranged in a three-dimensional vector. These
  features, previously used by \cite{Casetal2017} in analyzing several
similar fall data sets, are: $\log \left( \max_t SMV_t\right)$,
$\log\left( \min_t SMV_t\right)$, and
$\log\left( \max_t |SMV_t - SMV_{t-1}|\right) $.  Ultimately,
the resulting data set contained $154$ three-dimensional vectors
of extracted log-features.

\begin{figure}[t!]
\includegraphics[width=.90\textwidth]{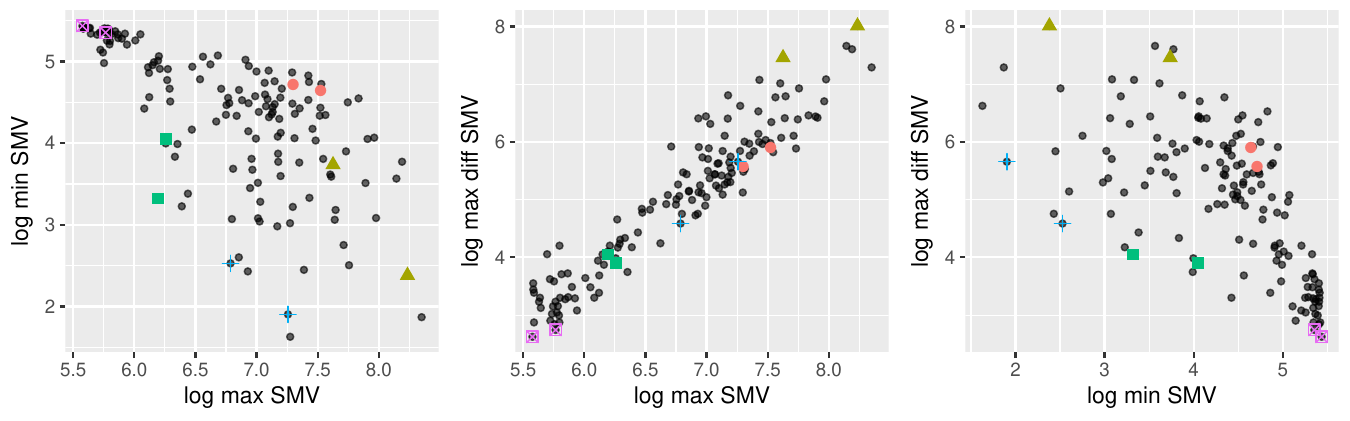} 
\caption{\label{fig::anchorpts_fall} The data and selected anchor
  points for the SisFall data example.}  
  \centering
\end{figure}

We fit a multivariate Gaussian mixture model with $k = 5$ components.
We selected the number of components based on the Bayesian information criterion (BIC)
%integrated completed
%likelihood (ICL) criterion \citep{Biernacki2000, Baudry_etal2010}
evaluated at MAP estimates of the exchangeable model parameters.  We
specified a $N_3(\vv{\mu},\vv{\Sigma}_j/\kappa)$ prior on
$\vv{\theta}_j$ with $\vv{\mu} = \bar{\vv{Y}}$, the sample mean vector
of the data, and $\kappa = 0.01$. We specified a Wishart
$(\nu, \vv{A})$ distribution on $\vv{\Sigma}_j^{-1}$ with $\nu=6$
degrees of freedom and prior scale $\vv{A} = 5\vv{I}_3$, where
$\vv{I}_p$ denotes the $p \times p$ identity matrix.  Finally, we
specified a Dirichlet$(\vv{1}_3)$ prior for~$\vv{\eta}$. We used the
anchored EM algorithm to select two anchor points per component.  The
data and selected anchor points are shown in
Figure~\ref{fig::anchorpts_fall}.  The coefficient of
quasi-consistency was estimated as
$\widehat{\alpha}>0.9999$.  Qualitatively,
the selected anchor points identify well-separated sites on the
periphery of the data cloud, as we would expect in a location problem
by generalizing the intuition provided by
Proposition~\ref{prop.k2results}.  The high value of
$\widehat{\alpha}$ indicates that we can
expect our anchor points to produce high posterior concentration on
the true parameter values in large samples. We fit the model using a
Gibbs sampler with $200$ parallel chains.  We thinned the chains
after $5,000$ burn-in iterations to obtain $M=50,000$
samples from the posterior distribution.

\begin{figure}[tb]
\includegraphics[width=.98\textwidth]{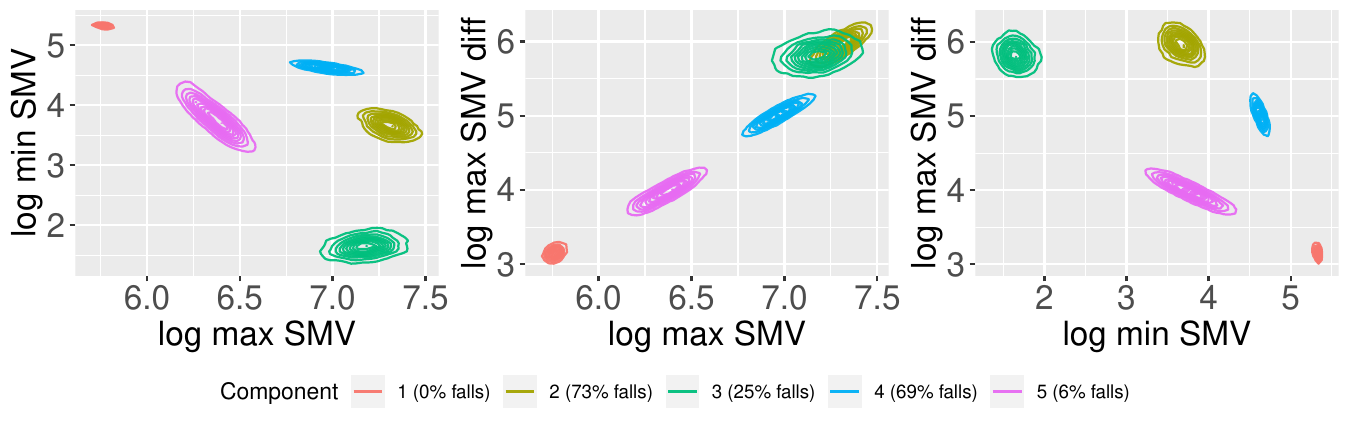}
\caption{\label{fig::fall_thetadist} 2D marginal density
  estimates of the 
  posterior distribution of $\vv{\theta}_j$ for the fall dataset. } 
\end{figure}

Posterior density estimates of $\vv{\theta}$ are shown in
Figure~\ref{fig::fall_thetadist}. 
Table~\ref{table::fall_allocations} lists the average posterior allocation
probabilities for selected activities, where each trial's
probability of allocation to component~$j$ is the relative frequency that $s_i^m =
j$, calculated from the Monte Carlo posterior samples of $\vv{s}^m$,
$m=1,\ldots,M$.  The table gives these probabilities averaged over the repeated trials for each
activity.
The legend of Figure~\ref{fig::fall_thetadist} 
%\remove{first row of Table~\ref{table::fall_thetaj}} 
also displays the proportion of falls among the observations
classified to each component, if each trial is classified to its
most probable component.

Component~1, whose mean is located in a far
corner of the posterior parameter space, is
characterized by low values of maximum SMV and high values of minimum
SMV throughout the trial. It is unsurprising that no activities classified to this
component are falls because falls are expected to be
associated with large changes in acceleration.  Component~5 describes activities with slightly higher values of maximum difference in SMV, and, unlike component~1, is estimated to contain a small number of falls.   Table~\ref{table::fall_allocations}
indicates that quick vertical movements, such as quickly sitting and standing (D10), are likely to be classified to
this component.  

Components~2 and~4 both exhibit high values of maximum SMV and maximum difference in SMV.  The majority of activities classified to these components are falls, with a few ADLs such as moving up and down stairs (D06), or trying to get up but
collapsing into a chair (D11).  The forward falls tend to be classified into component~2, whose SMV values are higher, while falls in other directions are classified into component~4. 
Component~3, like component~2, describes activities with very high values of maximum SMV and maximum difference, but unlike component~2 its average value of minimum SMV is low.   This component contains only 25\% falls, suggesting that minimum SMV is a feature that is able to distinguish ADLs from falls.  

% latex table generated in R 3.5.2 by xtable 1.8-3 package
% Fri May 15 16:36:10 2020
\begin{table}[ht]
	\small
	\centering
	\caption{\label{table::fall_allocations} Posterior allocation probabilities for selected activities in the SisFall dataset.}
	\begin{tabular}{p{10cm}rrrrr}
		\hline
		Activity &  \multicolumn{5}{l}{Component }\\
		& 1 & 2 & 3 & 4 & 5 \\ 
		\hline
%		D01 Walking slowly & 0.099 & 0.008 & 0.000 & 0.490 & 0.403 \\ 
%		D02 Walking quickly & 0.000 & 0.086 & 0.000 & 0.066 & 0.848 \\ 
		D03 Jogging slowly & 0.000 & 0.000 & 1.000 & 0.000 & 0.000 \\ 
%		D04 Jogging quickly & 0.000 & 0.011 & 0.989 & 0.000 & 0.000 \\ 
%		D05 Walking upstairs and downstairs slowly & 0.059 & 0.027 & 0.000 & 0.887 & 0.027 \\ 
		D06 Walking upstairs and downstairs quickly & 0.000 & 0.796 & 0.188 & 0.000 & 0.016 \\ 
		D07 Slowly sit in a half height chair, wait a moment, and up slowly  & 0.981 & 0.000 & 0.000 & 0.006 & 0.013 \\ 
	%	D08 Quickly sit in a half height chair, wait a moment, and up quickly & 0.000 & 0.019 & 0.001 & 0.024 & 0.956 \\ 
		D09 Slowly sit in a low height chair, wait a moment, and up slowly & 0.245 & 0.050 & 0.000 & 0.575 & 0.131 \\ 
		D10 Quickly sit in a low height chair, wait a moment, and up quickly & 0.001 & 0.042 & 0.001 & 0.148 & 0.808 \\ 
		D11 Sitting a moment, trying to get up, and collapse into a chair & 0.000 & 0.541 & 0.003 & 0.454 & 0.002 \\ 
%		D12 Sitting a moment, lying slowly, wait a moment, and sit again & 0.938 & 0.020 & 0.000 & 0.035 & 0.007 \\ 
%		D13 Sitting a moment, lying quickly, wait a moment, and sit again & 0.173 & 0.110 & 0.000 & 0.695 & 0.022 \\ 
%		D14 Being on one's back change to lateral position, wait a moment, and change to one's back & 0.942 & 0.001 & 0.000 & 0.038 & 0.019 \\ 
%		D15 Standing, slowly bending at knees, and getting up & 0.978 & 0.000 & 0.000 & 0.010 & 0.012 \\ 
%		D16 Standing, slowly bending without bending knees, and getting up & 0.986 & 0.000 & 0.000 & 0.008 & 0.005 \\ 
%		D17 Standing, get into a car, remain seated and get out of the car & 0.797 & 0.003 & 0.000 & 0.156 & 0.045 \\ 
		D18 Stumble while walking & 0.000 & 0.957 & 0.003 & 0.040 & 0.000 \\ 
		D19 Gently jump without falling (trying to reach a high object) & 0.000 & 0.308 & 0.033 & 0.000 & 0.658 \\ 
	%	F01 Fall forward while walking caused by a slip & 0.000 & 0.915 & 0.001 & 0.084 & 0.000 \\ 
		F02 Fall backward while walking caused by a slip & 0.000 & 0.302 & 0.001 & 0.697 & 0.000 \\ 
		%F03 Lateral fall while walking caused by a slip & 0.000 & 0.250 & 0.000 & 0.750 & 0.000 \\ 
		F04 Fall forward while walking caused by a trip & 0.000 & 0.938 & 0.001 & 0.061 & 0.000 \\ 
%		F05 Fall forward while jogging caused by a trip & 0.000 & 0.799 & 0.201 & 0.000 & 0.000 \\ 
%		F06 Vertical fall while walking caused by fainting & 0.000 & 0.058 & 0.000 & 0.942 & 0.000 \\ 
%		F07 Fall while walking, with use of hands in a table to dampen fall, caused by fainting & 0.000 & 0.211 & 0.001 & 0.788 & 0.000 \\ 
%		F08 Fall forward when trying to get up & 0.001 & 0.586 & 0.001 & 0.412 & 0.000 \\ 
		F09 Lateral fall when trying to get up & 0.000 & 0.074 & 0.000 & 0.926 & 0.000 \\ 
		F10 Fall forward when trying to sit down & 0.000 & 0.609 & 0.001 & 0.390 & 0.000 \\ 
%		F11 Fall backward when trying to sit down & 0.000 & 0.230 & 0.002 & 0.760 & 0.008 \\ 
%		F12 Lateral fall when trying to sit down & 0.000 & 0.239 & 0.001 & 0.759 & 0.000 \\ 
%		F13 Fall forward while sitting, caused by fainting or falling asleep & 0.014 & 0.726 & 0.001 & 0.124 & 0.135 \\ 
%		F14 Fall backward while sitting, caused by fainting or falling asleep & 0.001 & 0.527 & 0.003 & 0.467 & 0.002 \\ 
%		F15 Lateral fall while sitting, caused by fainting or falling asleep & 0.000 & 0.439 & 0.001 & 0.561 & 0.000 \\ 
		\hline
	\end{tabular}
\end{table}

 Table~\ref{table::fall_allocations} indicates that certain types of
 ADLs, such as sitting slowly (D07), 
 %, exhibit features that are distinct from the other types and 
 are unlikely to be confused with falls as indicated by their high
 probability of allocation to component~1.  The small maximum change in
  (log) acceleration associated with this component, is a feature that is
 likely to be highly predictive of certain ADLs.  Other ADLs, such as
 going upstairs quickly (D06), share the high-acceleration features
 that many falls exhibit.  The similarities between ADLs that involve
 fast movement and forward falls suggest that additional measurements, perhaps some including
 a directional component, may aid in better distinguishing falls in these difficult cases.
% This analysis provides insight into which ADLs may be easily confused with falls, with model parameters highlighting those features that certain ADLS share with falls. 

\vspace*{-8mm}
\section{Discussion} \label{section:discussion} \vspace*{-2mm} The
proposed anchored Bayesian mixture model offers a model-based
resolution to label-switching that eliminates prior and posterior
exchangeability without imposing highly restrictive identifiability
constraints.  In Section~\ref{subsection:large_sample} we introduced a notion of quasi-consistency
which guarantees that, in large samples, a well-specified anchor model places high posterior probability on 
one relabeling of the true component-specific parameters. 
Our anchored EM strategy for selecting optimal anchor points requires 
several pre-processing steps, but
eliminates the need for post-processing of MCMC samples, often resulting in computational savings.  A carefully-specified anchor model will produce
component-specific parameter estimates that reflect homogeneous
subgroups in the population and arise directly from the specified
model.  

The examples presented in this paper have demonstrated that one or two anchor points per component are often sufficient to eliminate posterior multimodality and produce accurate parameter estimates.  The optimal number of anchor points may, in some situations, be higher, and may depend on the method used to  select the points.  Future work will investigate this question in a variety of univariate and multivariate settings.   

%Several existing model-based approaches prevent
%posterior label ambiguity by specifying non-exchangeable priors on the 
%mixture proportions \citep{mena2015} or repulsive priors that favor component-specific parameters that are well-separated \citep{PetRaoDun2012, XuMuller2016, XieXu2019},
%).  The anchor model joins these model-based approaches as an easy-to-implement modification of the simple exchangeable model that can be used with conditional conjugate priors. 

%It is straightforward to implement and appropriate for many
%applications where component-specific inference is a central goal of
%the analysis.  
\paragraph{Non-Gaussian components.} 
The anchor model methodology is readily applicable to non-Gaussian
component distributions.  The model properties that we presented in
this paper do not, in general, rely on Gaussian components.  Anchoring
always eliminates the model's posterior exchangeability and a minimal
number of anchor points will typically produce distinct distributions
for the $\gamma_j$, as outlined in
Proposition~\ref{prop:proper_prior_equiv}, when the component
distributions are continuous in $\vv{y}$ and $\vv{\gamma}$.
Proposition~\ref{prop:increasing_marg_like}, which stated that the
anchor model's fit improves with the addition of more anchor points,
is also true for non-Gaussian mixtures under fairly general
  conditions.  The asymptotic result in
Section~\ref{subsection:large_sample} does depend on several
regularity conditions on the component likelihoods and priors, which
may not hold for all choices of component distributions.
%The expressions $m(\vv{y}|\vv{s})$ that indicate conditional goodness of fit are
%specific to the Gaussian mixture, and so the heuristic notions about
%the shapes of anchor models that fit well may differ when other models
%are considered. 
The anchored EM algorithm of
Section~\ref{section::model_specification} can be implemented in
models from other families.  Our own applied work \citep{kun:per:2019} has demonstrated the use of this strategy in a Gaussian mixture of regressions model.
Different component distributions will motivate new approaches to specifying anchor points and
these are interesting directions of future research.

\paragraph{Other extensions.} 
Future work will explore ways to quantify the sensitivity of analyses
to changes in the specification of the anchor points and the number of
anchor points.  An interesting feature of anchor models is that they
admit the possibility of using improper priors for the
component-specific parameters.  Sensitivity to proper priors is a
well-known issue in mixture modeling \citep{RicGre1997, Fru2006} but
non-informative priors are difficult to derive because improper priors
typically lead to improper posteriors \citep{GrazianRobert2018}.  An
anchor model can restrict the space of latent allocations to prevent
empty components, similarly to the methods proposed by
\cite{DieboltRobert1994} and~\cite{wasserman2000}, and this
possibility allows investigation of the model's behavior under a wider
class of priors.

From an applied modeling perspective, we plan to extend the anchored
mixture methodology to the case of hierarchical mixture models fit to
grouped data collected on many experimental units, as in the case, for
example, of the entire SisFall data set.  Assuming a mixture model
with a fixed number of components for the data collected on each of
the experimental units, with component specific parameters tied
together in a hierarchical structure, several challenging modeling
questions will need an answer.  Decisions will have to be made
concerning the number of components needed to describe the data for
each experimental unit.  A simple approach would employ the same
number (possibly random) of components for each subject.  A more refined approach would
allow for varying numbers of components across units.  With specific
regard to the anchored methodology, we plan to investigate different
approaches to the specification of the anchor points. 
These include selecting anchor points using
independent fits to the data for each unit and strategies that account for existing dependencies in the data.  We will also consider approaches where only a subset of the units will have anchored observations.

\vspace*{-9mm}
\section*{Acknowledgments}
\vspace*{-4mm} This paper is based on results presented in the first
author's Ph.D.~dissertation \citep{Kunkel2018}.  The authors' work was
partially supported by the National Science Foundation under Grant
No.~SES-1424481 and~No.~SES-1921523. The authors thank Steven
MacEachern, Bettina Gr\"{u}n, and two anonymous reviewers for
their helpful comments.

%\section*{References}

%%\setlength{\bibsep}{0pt}\setlength{\bibhang}{4pt}
\bibliographystyle{apalike}
%%\bibliography{mixturemodels_mario}
%\vspace{-13mm}
\bibliography{mixturemodels}

\clearpage

\appendix

\renewcommand{\thetable}{A\arabic{table}}
\renewcommand{\thefigure}{A\arabic{figure}}

\begin{center}
	\section*{Anchored Bayesian Gaussian Mixture Models}
	\textbf{Deborah Kunkel and Mario Peruggia}
	
	\vspace{.2cm}
	\section*{Appendix}
\end{center}

\section{Proofs of the propositions}
This section presents proofs of Propositions~1 and 3-6.  When
necessary, references to expressions in the main manuscript will be
preceded by M, so that, for example,
\manuscriptref{GMM_latentallocations} refers to
Equation~\eqref{GMM_latentallocations}
in the main manuscript.
%\begin{prop}
%Consider an anchor model $A =\{A_1,\ldots,A_k \}$ with $m$
%  anchor points.  Let $\mathscr{S}^A$ be the subset of allocations
%that has nonzero probability under $A$.
%Then $\mathscr{S}^A$ contains $k^{n-m}$ elements and $A$ assigns
%probability zero to every allocation
%that satisfies $s_i = s_i'$, for some $i \in
%  A_j$ and $i' \in A_{j'}$, $j \neq j'$. $\quad$
%\end{prop}
%\paragraph{Proof.}
% Let $\mathcal{S}^A$ be the subset of allocations that has nonzero probability under the anchor model $A$.  Then $\mathcal{S}^A$ contains $k^{n-m}$ elements and $A$ assigns probability of zero to every partition that satisfies $s_i \neq s_i'$, for all $i \in A_j$ and $i' \in A_{j'}$, $j \neq j'$.  
%
%The size of $\mathcal{S}^A$ can be seen by observing that each possible allocation $\vv{s} \in \mathcal{S}^A$ has $m$ fixed elements corresponding to the anchor points, and the remaining $n-m$ elements in $\vv{s}$ can take all values in $1,\ldots,k$.
%The second statement follows from the definition of the anchor model.
%

\begin{customprop}{\ref{prop:unique_labeling}}
	\label{A.prop:unique_labeling}   The following two statements hold under conditions C.1 and C.2.
	\begin{enumerate}
		\item    	An anchor model $A
		=\{A_1,\ldots,A_k \}$ imposes a unique labeling on each partition 
		that has nonzero probability if and only if $A_1,\ldots,A_{k-1}$ are
		non-empty; that is, $k_0\geq k-1$. 
		\item  For any $j\leq k_0$, $j\neq j\prime$, the marginal
		posterior density of $\vv{\gamma}_j$ is distinct from the marginal
		posterior density of $\vv{\gamma}_{j\prime}$ with probability~1.
	\end{enumerate}
	
\end{customprop}
\paragraph{Proof of part a.} Suppose $A_{k-1}$ and $A_k$ are empty so that at
least two components have no points anchored to them.  Choose an
allocation $\vv{s}^*$ from $\mathscr{S}^A$ such that $\vv{s}^*$ has at
least one element equal to $k-1$, so that $\vv{s}^*$ has induced a
partition of the data under which one group is labeled $k-1$. The set
$\mathscr{S}^A$ also contains the allocation obtained by permuting the
label $k$ and $k-1$ in $\vv{s}^*$, which induces the same partition
but a different labeling.  In contrast, if $A_1,\ldots,A_{k-1}$ each
contain at least one point, any allocation from $\mathscr{S}^A$
induces a partition that cannot be relabeled without relabeling an
anchor point.   

\paragraph{Proof of part b.} Assume that component $j$ has at least one anchor point; that is, $A_j \neq \emptyset$, and that $\vv{x}_h$ denotes the anchor points for component~$h$. 
Use $\vv{\gamma}_{-(h)}$ to denote $\vv{\gamma}=(\vv{\gamma}_1,\ldots,\vv{\gamma}_{h-1},\vv{\gamma}_{h+1},\ldots,\vv{\gamma}_k)$ and $\Gamma$ to denote the parameter space of $\vv{\gamma}_h$.  The posterior density of $\vv{\gamma}_j$, denoted by $p_j(\vv{\gamma}_j|\vv{y})$, satisfies
\small
\begin{align}
p_j(\vv{\gamma}_j|\vv{y}) \propto &
\int_{\vv{\gamma}_{-(j)}}\int_{\vv{\eta}} p(\vv{\gamma}_1, \ldots, \vv{\gamma}_k, \vv{\eta}, \vv{y})d\vv{\eta}d\vv{\gamma_{-(j)}} \\
= & \int_{\vv{\gamma}_{-(j)}}\int_{\vv{\eta}}\pi(\vv{\eta})\prod_{h=1}^{k_0}\phi_p(\vv{x}_h;\vv{\gamma}_h)\prod_{h=1}^k\pi(\vv{\gamma}_h)\prod_{i \not\in A}\sum_{h=1}^k\eta_h\phi_p(y_i;\vv{\gamma}_h) d\vv{\eta} d\vv{\gamma}_{-(j)}. \label{postgammaj_1}
\end{align}
\paragraph{Case 1:  $A_{j\prime} \neq \emptyset$.}  
Define $c(\vv{\gamma}_j,\vv{\gamma}_{j\prime})$ to be equal to the following function:
\begin{align}
c(\vv{\gamma}_j,\vv{\gamma}_{j\prime}) &=\int_{\vv{\gamma}_{-(j, j\prime)}}\prod_{h \neq j, j\prime}^{k_0}\phi_p(\vv{x}_h;\vv{\gamma}_h)\prod_{h \neq j, j\prime}^k\pi(\vv{\gamma}_h) \int_{\vv{\eta}}\pi(\vv{\eta})\prod_{i \not\in A}\sum_{h=1}^k\eta_h\phi_p(y_i;\vv{\gamma}_h)d\vv{\eta} d\vv{\gamma}_{-(j,j\prime)}, 
\end{align}
noting that the label-invariance of the mixture likelihood gives 
\begin{align}c(\vv{u}, \vv{w}) = c(\vv{w}, \vv{u}). \label{c_uw}
\end{align}
The expression~\eqref{postgammaj_1} can be written as
\begin{align}
p_j(\vv{\gamma}_j|\vv{y}) \propto & \pi(\vv{\gamma}_{j})\phi_p(\vv{x}_{j};\vv{\gamma}_{j})\int_{\vv{w}}\pi(\vv{w})\phi_p(\vv{x}_{j\prime};\vv{w})c(\vv{w}, \vv{\gamma}_{j})d\vv{w}.
\end{align}
Further, define 
\begin{align}
c_2(\vv{\gamma}_j, \vv{x}_{j\prime}) &= \int_{\vv{u}}\pi(\vv{u})\phi_p(\vv{x}_{j\prime};\vv{u})c(\vv{\gamma}_j, \vv{u})d\vv{u}, \end{align}
yielding
\begin{align}
p_j(\vv{\gamma}_j|\vv{y}) \propto & \pi(\vv{\gamma}_{j})\phi_p(\vv{x}_{j};\vv{\gamma}_{j})c_2(\vv{\gamma}_j, \vv{x}_{j\prime}).
\end{align}
For component $j\prime$, we have
\begin{align}
p_{j\prime}(\vv{\gamma}_{j\prime}|\vv{y}) \propto & \pi(\vv{\gamma}_{j\prime})\phi_p(\vv{x}_{j\prime};\vv{\gamma}_{j\prime})\int_{\vv{w}}\pi(\vv{w})\phi_p(\vv{x}_{j};\vv{w})c(\vv{w}, \vv{\gamma}_{j\prime})d\vv{w} \\
&= \pi(\vv{\gamma}_{j\prime})\phi_p(\vv{x}_{j\prime};\vv{\gamma}_{j\prime})c_2(\vv{\gamma}_{j\prime},\vv{x}_j),
\end{align} 
where the second equality follows from~\eqref{c_uw}.
Assumption C.2 ensures that there is some open set $W \subset \Gamma$ such that $\pi(\vv{w})>0$ for all $\vv{w}\in W$.  Then
for $\vv{w} \in W$, we have the posterior density of $\gamma_j$ satisfying	
\begin{align}
p_j(\vv{w}|\vv{y}) \propto & \pi(\vv{w})\phi_p(\vv{x}_{j};\vv{w})c_2(\vv{w}, \vv{x}_{j\prime}),
\end{align}
while the posterior density of $\vv{\gamma}_{j\prime}$ satisfies
\begin{align}
p_{j\prime}(\vv{w}|\vv{y}) \propto & \pi(\vv{w})\phi_p(\vv{x}_{j\prime};\vv{w})c_2(\vv{w}, \vv{x}_{j}).
\end{align}
With probability~1, $\vv{x}_j \neq \vv{x}_{j\prime}$, and so $p_{j}(\vv{w}|\vv{y}) \neq p_{j\prime}(\vv{w}|\vv{y})$ for all $\vv{w} \in W$.

\paragraph{Case 2.}
If $A_{j\prime} =\emptyset$, then 
\begin{align}
p_j(\vv{\gamma}_j|\vv{y}) \propto & \pi(\vv{\gamma}_{j})\phi_p(\vv{x}_{j};\vv{\gamma}_{j})\int_{\vv{w}}\pi(\vv{w})c(\vv{w}, \vv{\gamma}_{j})d\vv{w}
\end{align}
while
\begin{align}
p_{j\prime}(\vv{\gamma}_{j\prime}|\vv{y}) \propto & %\pi(\vv{\gamma}_{j\prime})\int_{\vv{\gamma}_{j}}\pi(\vv{\gamma}_{j})\phi_p(\vv{x}_{j};\vv{\gamma}_{j})c(\vv{\gamma}_{j}, \vv{\gamma}_{j\prime})d\vv{\gamma_{j}} \\
\pi(\vv{\gamma}_{j\prime})c_2(\vv{\gamma}_{j\prime},\vv{x}_j).
\end{align} 
For $\vv{w} \in W$, is it clear that $p_{j}(\vv{w}|\vv{y}) \neq 	p_{j\prime}(\vv{w}|\vv{y})$.

\begin{customprop}{\ref{pq_expression_prop}}
	The $q$th element of $P_{\vv{x}}(\vv{\gamma}_0)$, $p_q$, is equal to
	\begin{align} \label{pq_expression}
	\prod_{j=1}^{k_0}\phi_p\left(\vv{x}_j; \vv{\gamma}_{0\rho_q(j)}\right)\left/
	\sum_{h=1}^{k!}\prod_{j=1}^{k_0}\phi_p\left(\vv{x}_j;\vv{\gamma}_{0\rho_h(j)}\right)\right..
	\end{align}
\end{customprop}

\paragraph{Proof:} 
The data dependent prior of $(\vv{\gamma},\vv{\eta})$ under model $A$ is proportional to $$\pi(\vv{\eta})\prod_{j=1}^{k_0}p(\vv{\gamma}_j|\vv{x}_j)\prod_{j=k_0+1}^{k}\pi(\vv{\gamma}_j).$$  

The probability of the $q$th class label under the anchor model is 
\begin{align} %\label{A.prob_qth}
p_q & = \frac{\pi(\rho_q(\vv{\eta}_0))\prod_{j=1}^{k_0}p(\vv{\gamma}_{0\rho_q(j)}|\vv{x}_{\rho_q(j)})\prod_{j=k_0+1}^{k}\pi(\vv{\gamma}_{0\rho_q(j)})}{\sum_{h=1}^{k!}\pi(\rho_h(\vv{\eta}_0))\prod_{j=1}^{k_0}p(\vv{\gamma}_{0\rho_h(j)}|\vv{x}_j)\prod_{j=k_0+1}^{k}\pi(\vv{\gamma}_{0\rho_h(j)})} \\
&=\frac{\prod_{j=1}^{k_0}p(\vv{\gamma}_{0j}|\vv{x}_j)\prod_{j=k_0+1}^{k}\pi(\vv{\gamma}_{0\rho_q(j)})}{\sum_{h=1}^{k!}\prod_{j=1}^{k_0}p(\vv{\gamma}_{0\rho_h(j)}|\vv{x}_j)\prod_{j=k_0+1}^{k}\pi(\vv{\gamma}_{0\rho_h(j)})}, \label{pq1}
\end{align}
where the terms $\pi(\rho_q(\vv{\eta}_0))$ are invariant to label permutation and are factored out in~\eqref{pq1}.  For any $q$, we have 
\begin{align}  \label{A.posterior_anchorpts}
\prod_{j=1}^{k_0}p(\vv{\gamma}_{0\rho_q(j)})|\vv{x}_j)&=  \prod_{j=1}^{k_0}\pi(\vv{\gamma}_{0\rho_q(j)}))\phi_p(\vv{x}_j; \vv{\gamma}_{0\rho_q(j)}))C_{j}^{-1}.
\end{align}
As in Section~\ref{subsection::Anchor_informativeprior} of the manuscript, $C_j$ is defined to be  
\begin{align*}
C_j  = \int\pi(\vv{w})\phi_p(\vv{x}_j; \vv{w})d\vv{w}
\end{align*}
and does not depend on the permutation of the label of $\vv{\gamma}_j$. 
Further,  $\prod_{j=1}^k\pi(\vv{\gamma}_{\rho_q(j)})$ is  equal to $\prod_{j=1}^k\pi(\vv{\gamma}_{\rho_{q\prime}(j)}) $ for any $q$ and $q\prime$. The only term in~\eqref{pq1} that depends on the $q$th permutation is $\prod_{j=1}^{k_0}\phi_p(\vv{x}_j;\vv{\gamma}_{\rho_q(j)})$.
Thus the $q$th element of $P_{\vv{x}}(\vv{\gamma}_0)$ is equal to 
\begin{align*}
p_q & =  \frac{\prod_{j=1}^{k_0}\phi_p(\vv{x}_j;
	\vv{\gamma}_{0\rho_q(j)})}{\sum_{h=1}^{k!}\prod_{j=1}^{k_0}\phi_p(\vv{x}_j;\vv{\gamma}_{0\rho_h(j)})}
\end{align*} 
for $q=1,\ldots,k!$.

\begin{customprop}{\ref{prop.k2results}}
	Suppose that $p=1$, $k=2$, and that  $m_j = m$ observations (with $1\leq m \leq n/2$) are to be anchored to component $j$, $j=1,2$.
	The following results hold:
	\begin{enumerate}
		\item  If $\sigma_1^2 =  \sigma_2^2= \sigma^2$ and $\theta_1 < \theta_2$, then the optimal anchoring sets
		$\vv{x}_1 = ( y_{(1)},\ldots, y_{(m)})$, $\vv{x}_2 =( y_{(n-m+1)},
		y_{(n)} )$, where $y_{(l)}$ denotes the $l$th order statistic.  %If $m_1 = m_2 = 1$, the minimum and maximum values should be chosen as anchor points. 
		\item  If $\theta_1 = \theta_2 = \theta$ and $\sigma_1^2 < \sigma_2^2$,
		then the optimal anchoring sets $\vv{x}_1$ equal to the points
		that minimize 
		$\sum_{i=1}^{m} (y_i - \theta)^2$
		and $\vv{x}_2$ equal to the
		points that maximize 
		$\sum_{i=1}^{m} (y_i - \theta)^2$. 
	\end{enumerate}
\end{customprop}

\paragraph{Proof.} When $k=2$, $P_{\vv{x}}$ has only two elements and maximizing $p_1$ will minimize its entropy. Thus, it is sufficient to maximize the ratio $p_1/p_2$.  In the location problem (case 1), $p_1/p_2$ equals
\begin{align*}
\frac{\phi(\vv{x}_1;\theta_1,\sigma^2)\phi(\vv{x}_2;\theta_2,\sigma^2)}{\phi(\vv{x}_1;\theta_2,\sigma^2)\phi(\vv{x}_2;\theta_1,\sigma^2)} &= \exp\left(\frac{m}{\sigma^2} (\theta_{2}-\theta_{1})(\bar{x}_2-\bar{x}_1)\right),
\end{align*}
where $\bar{x}_1$ and $\bar{x}_2$ are the sample means of $\vv{x}_1$ and $\vv{x}_2$.  Because we assume $\theta_1 < \theta_2$, this expression is increasing in $(\bar{x}_2-\bar{x}_1)$. 
In case 2, $p_1/p_2$ is equal to
\begin{align*}
\begin{split}
\frac{\phi(\vv{x}_1;\theta, \sigma_1^2)\phi(\vv{x}_2;\theta, \sigma_2^2)}{\phi(\vv{x}_1;\theta, \sigma_2^2)\phi(\vv{x}_2;\theta, \sigma_1^2)} &=\exp\left(\left(\frac{1}{2\sigma_1^2}-\frac{1}{2\sigma_2^2}\right)\left(\sum_{i=1}^{m}(x_{2i}-\theta)^2 - \sum_{i=1}^{m}(x_{1i}-\theta)^2\right)\right). %\label{A.p1p2_scale}
\end{split}
\end{align*}
Because $\sigma_1^2 < \sigma_2^2$, the ratio is increasing in
$\sum_{i=1}^{m}(x_{2i}-\theta)^2 $ and decreasing in
$\sum_{i=1}^{m}(x_{1i}-\theta)^2$.  This concludes the proof.
\vskip 15mm

\begin{customprop}{\ref{prop:increasing_marg_like}}
	\label{Aprop:increasing_marg_like} Assume that $\eta_j =
	1/k$, $j=1,\ldots,k$.  
	Let $A_*^1,\ldots, A_*^n$ be a sequence of
	anchor models where $A_*^m$ has the highest marginal likelihood
	among all anchor models with $m$ anchor points. The marginal
	likelihoods of the models satisfy
	$m_{A_*^{1}}(\vv{y}) \leq \ldots \leq m_{A_*^{n}}(\vv{y})$.
\end{customprop}
The proof of Proposition~\ref{prop:increasing_marg_like} below relies on the result stated in the following 
lemma.
\begin{lemma} \label{nested_A}
	For any anchor model $A^{m}$, $m\geq 1$, it is possible to form a sequence of anchor models $A^{m}, \ldots, A^n$ satisfying $A^m \subset A^{m+1} \subset \ldots A^n$ such that $m_{A^m}(\vv{y}) \leq \ldots \leq m_{A^n}(\vv{y})$.
\end{lemma}
%\paragraph{Proof of Lemma~\ref{nested_A}.}
\paragraph{Proof.}
Given an anchor model $A^m$
with $m$ anchor points and some $i \notin A^m$, let $A^{m+1}_{i,j}$
denote the anchor model obtained from $A^m$ by adding the additional
anchor point with index $i$ to component~$j$.  We will show that for
some $j$, the marginal likelihood for $A^{m+1}_{i,j}$ is greater than
or equal to that of $A^{m}$.

Let $\mathscr{S}^{A^{m}}$ denote the restricted set of allocation vectors for the anchor model $A^m$.  It is possible to write $\mathscr{S}^{A^m}$ as $\cup_{j=1}^k \mathscr{S}^{A^{m+1}_{i,j}}$ for any $i$. %, where $\mathscr{S}^{A^{m+1}_{i,j}} = \{ \vv{s}: s_i = j; \vv{s} \in \mathscr{S}^{A^m}\}$.   the set $\mathscr{S}^{A^{m+1}_{i,j}}$ contains the allocation vectors for 
The marginal likelihood for $A^m$  can be written as  
\begin{align}
m_{A^m}(\vv{y}) &= \sum_{\mathscr{S}^{A^{m}}} m(\vv{y}|\vv{s})p_{A^{m}}(\vv{s}) \nonumber \\ 
& = \frac{1}{k^{n-m}}\sum_{\mathscr{S}^{A^{m}}} m(\vv{y}|\vv{s}) \label{pa_simplified} \\ 
& = \frac{1}{k} \sum_{j=1}^k \frac{1}{k^{n-m-1}} \sum_{\mathscr{S}^{A^{m+1}_{i,j}}} m(\vv{y}|\vv{s}). \label{sum_sij}
\end{align}
Expression~\eqref{pa_simplified} follows from the fact that $p_{A^m}(\vv{s}) = k^{-(n-m)}$ for all $\vv{s} \in \mathscr{S}^{A^{m}}$ under the assumption that $\eta_j = k^{-1}$, $j=1,\ldots,k$.  Expression~\eqref{sum_sij} shows that the marginal likelihood for $A^m$ is the average of the marginal likelihoods for $A^{m+1}_{i,j}$, $j=1,\ldots,k$.  Thus,
\begin{align}
m_{A^m}(\vv{y}) & =\frac{1}{k} \sum_{j=1}^k \frac{1}{k^{n-m-1}} \sum_{\mathscr{S}^{A^{m+1}_{i,j}}} m(\vv{y}|\vv{s}) \nonumber \\
& \leq  \max_j \frac{1}{k^{n-m-1}} \sum_{\mathscr{S}^{A^{m+1}_{i,j}}} m(\vv{y}|\vv{s}) \nonumber \\
&= m_{A^m_{i,j^*}}(\vv{y}),
\end{align}
where $j^*$ is the component to which observation $i$ can be anchored to maximize the marginal likelihood.  Thus $m_{A^m}(\vv{y}) \leq m_{A^{m+1}_{i,j^*}}(\vv{y})$ for any $m<n$, proving Lemma~\ref{nested_A}. 

\begin{customprop}{\ref{prop:increasing_marg_like}}
	Assume that $\eta_j = 1/k$, $j=1,\ldots,k$. 
	Let $A_*^1,\ldots, A_*^n$ be a sequence of
	anchor models where $A_*^m$ has the highest marginal likelihood
	among all anchor models with $m$ anchor points. The marginal
	likelihoods of the models satisfy
	$m_{A_*^{1}}(\vv{y}) \leq \ldots \leq m_{A_*^{n}}(\vv{y})$.
\end{customprop} 

\paragraph{Proof.}
Fix $m$ and let $i$ be the index of an observation not anchored
in  $A_*^m$.  Then, 
\[
m_{A_*^{m}}(\vv{y}) \leq m_{A_{i,j^*}^{m+1}}(\vv{y}) \leq m_{A_*^{m+1}}(\vv{y}),
\]
where the first inequality follows from Lemma~\ref{nested_A} and the
second from the definition of $A_*^{m+1}$.

% \paragraph{Proof.}  The result follows directly from Lemma~\ref{nested_A}: 
% \begin{align}
% m_{A_*^{m}}(\vv{y}) &=  \frac{1}{k^{n-m}}\sum_{\mathscr{S}^{A_*^m}} m(\vv{y}|\vv{s}) \nonumber \\
% & = \frac{1}{k} \sum_{j=1}^k \left( \frac{1}{k^{n-m-1}} \sum_{\mathscr{S}^{A^{m+1}_{i,j}}} m(\vv{y}|\vv{s}) \right) \nonumber \\
% &\leq m_{A_{i,j^*}^{m+1}}(\vv{y}) \nonumber \\
% %& \sum_{\mathscr{S}^{A^{m+1}_*}} m(\vv{y}|\vv{s})p_{A_{*}^{m+1}}(\vv{s})\\
% %&= \frac{1}{k^{n-(m+1)}}\left(\frac{1}{k} \sum_{j=1}^k \sum_{\mathscr{S}^{A^{m+1}}(i,j)} m(\vv{y}|\vv{s}) \right) \\
% %\label{A.step_three} & \leq \frac{1}{k^{n-(m+1)}} \left( \sum_{\mathscr{S}^{A_*^{m+1}}} m(\vv{y}|\vv{s})\right) \\ 
% &\leq m_{A_*^{m+1}}(\vv{y}).
% \end{align}

\newpage
\begin{customprop}{\ref{prop:anchorq}} 
	Let $q$ be the posterior
	distribution of the allocations under an anchor model, subject to
	the restrictions in \manuscriptref{anchored_qs}.    The KL divergence of
	$q_*$ from $q$, evaluated at a fixed value of $\vv{\gamma}$, is
	minimized when the sets $A_j$ are chosen to maximize
	$\sum_{j=1}^{k_0}\sum_{i \in A_j}r_{ij}$ and when $\widetilde{r}_{ij} = r_{ij}$ for all $i \not \in A$.
\end{customprop}

\paragraph{Proof.} For any distributions $q$ and $q_*$, and defining
$x\log(x)=0$, the KL-divergence of $q_*$ from $q$
is equal to 
\begin{align*}
D_{KL}(q||q_*) &= \sum_{\vv{s}}q(\vv{s})\log\left(\frac{q(\vv{s})}{q_*(\vv{s})}\right) \\
&=\sum_{\vv{s}}\left[ \prod_{l=1}^nq(s_l)\right]\log\left(\frac{\prod_{i=1}^nq(s_i)}{\prod_{i=1}^nq_*(s_i)}\right)
\\
&=\sum_{\vv{s}}\left[ \prod_{l=1}^nq(s_l)\right]\sum_{i=1}^n\left(\log\left(\frac{q(s_i)}{q_*(s_i)}\right)\right)
\\
&=\sum_{s_1=1}^k\ldots\sum_{s_n = 1}^{k}\left[ \prod_{l=1}^nq(s_l)\right]\sum_{i=1}^n\left(\log\left(\frac{q(s_i)}{q_*(s_i)}\right)\right)
\\
&=\sum_{s_1=1}^k\ldots\sum_{s_{n-1}=1}^k\left[ \prod_{l=1}^{n-1}q(s_l)\right]\sum_{s_n = 1}^{k}q(s_n)\sum_{i=1}^n\left(\log\left(\frac{q(s_i)}{q_*(s_i)}\right)\right)
\\
\begin{split}
&=\sum_{s_1=1}^k\ldots\sum_{s_{n-1}=1}^k\left[ \prod_{l=1}^{n-1}q(s_l)\right]\\
& \;\;\;\;\;\;\;\;\;\;\left[\sum_{s_n = 1}^{k}q(s_n) \sum_{i=1}^{n-1}\left(\log\left(\frac{q(s_i)}{q_*(s_i)}\right)\right)+ \sum_{s_n = 1}^{k}q(s_n)\left(\log\left(\frac{q(s_n)}{q_*(s_n)}\right)\right)\right]   
\end{split}
%  \\
\end{align*}
\begin{align*}
&=\sum_{s_1=1}^k\ldots\sum_{s_{n-1}=1}^k\left[ \prod_{l=1}^{n-1}q(s_l)\right]\left[ \sum_{i=1}^{n-1}\left(\log\left(\frac{q(s_i)}{q_*(s_i)}\right)\right)+ \sum_{s_n = 1}^{k}q(s_n)\left(\log\left(\frac{q(s_n)}{q_*(s_n)}\right)\right)\right]   
\\
\nonumber & \vdots \\ 
&= \sum_{i=1}^n\sum_{s_i=1}^kq(s_i)\log\left(\frac{q(s_i)}{q_*(s_i)}\right).
\end{align*} 
%Because $q(s_i)=q^*(s_i)=t_{ij}$ for $i \notin A$, this simplifies to 
Substituting the definition of $q_*$ in~\manuscriptref{posterior allocation distribution} and the restrictions on $q$ from~\manuscriptref{anchored_qs} yields
\begin{align}
D_{KL}(q||q_*) &=\sum_{i=1}^n\sum_{j=1}^kq(s_i=j)\log\left(\frac{q(s_i=j)}{r_{ij}}\right) \nonumber \\
&=\sum_{i\in A_j}\sum_{j=1}^kI(s_i = j)\log\left(\frac{I(s_i = j)}{r_{ij}}\right) + \sum_{i\notin A}\sum_{j=1}^k\widetilde{r}_{ij}\log\left(\frac{\widetilde{r}_{ij}}{r_{ij}}\right) \nonumber \\
%&= -\sum_{j=1}^k\sum_{i \in A_j} \log\left(q^*(s_i=j|y_i,\vv{\gamma})\right) \\
% &= \sum_{j=1}^k\sum_{i \in A_j} \log\left(\frac{\sum_{l=1}^k\eta_lf(y_i|\gamma_l)}{\eta_jf(y_i|\gamma_j)}\right) \\
&= -\sum_{j=1}^k\sum_{i \in A_j}\log\left(r_{ij}\right) + \sum_{j=1}^k\sum_{i\notin A}\widetilde{r}_{ij}\log\left(\frac{\widetilde{r}_{ij}}{r_{ij}}\right). \label{A.DKL1}
\end{align} 
The second term in~\eqref{A.DKL1} is
non-negative and can be made equal to zero by setting
$\widetilde{r}_{ij} = r_{ij}$ for $i \notin A$. The first term decreases toward zero as $r_{ij} \rightarrow 1$ for $i \in A$.  Subject to the restriction that $A_j \cap A_{j^\prime} = \emptyset$ and $|A_j| = m_j$, the $q$ that minimizes
$D_{KL}(q||q_*)$ is
\begin{align*} \label{A.anchored_qs_sol}
q(S_i = j) &= \begin{cases} r_{ij} \quad & i \notin A \\
1 \quad & i \in A_j \\ 
0 \quad & i \in A_{j'}, \quad j' \neq j, \end{cases}
\end{align*}
where the $A_j$ are selected to maximize $\sum_{j=1}^k\sum_{i \in A_j}r_{ij}$. 
The sets $A_j$ will contain the $m_j$ observations with the
highest value(s) of $r_{ij}$, for $j=1,\ldots,k_0$, if there are no observations that fit this criterion for more than one value of $j$.

\newpage

\section{Simulation study}
Here, we describe in detail the simulation
studies referenced in Section~\ref{sec:no_of_points} of the main
manuscript, evaluating how the number of anchor points affects
out-of-sample predictions from the anchor model.

\subsection{Simulation 1.}
In the first simulation study, we simulated data from a mixture with $k=2$
mixture components to assess the relationship between goodness of fit
and predictive performance of the anchor model.  For each data set we
fit a univariate location model assuming $\sigma_1 = \sigma_2=1$ and
$\eta_1 = \eta_2 = 0.5$, using independent $N(0,25)$ prior
distributions on the component means $\theta_1$ and $\theta_2$.  We
generated data from a Gaussian mixture with means $\theta_1 = 0$,
$\theta_2 = \delta$ and standard deviations $\sigma_1 = 1$ and
$\sigma_2 = \sigma$.  We considered several values of $\delta$ to
assess the effect of separation among the mixture components and
several values of $\sigma$ to assess the effect of model
misspecification.  For each combination of $\delta$ and $\sigma$, we
generated 1,000 small data sets, $\vv{y}_{j,(\delta,\sigma)}$,
$j = 1, \ldots, 1,000$, of size $n=10$.  To each of these data sets,
we fit nine anchor models
$A_{j,(\delta,\sigma)}^2,\ldots,A_{j,(\delta,\sigma)}^{10}$, having
$m=2,\ldots,10$ anchor points such that
$A_{j,(\delta,\sigma)}^m$ has the highest marginal likelihood among
anchor models with $m$ anchor points, subject to the additional
requirement that at least one anchor point be assigned to each of
the two components.  Note that, to block out uninteresting sources
of variation, we used a common master batch of 1,000 data sets sampled
from a model with standard normal mixture components and obtained the
1,000 data sets for each $(\delta,\sigma)$ pair through appropriate
rescaling and translation of the observations.

We assessed the out-of-sample predictive performance of each
model in terms of its expected log pointwise predictive density
(ELPPD) \citep{Gelman2014} as follows.  For a given simulated data set,
$\vv{y}_{j,(\delta,\sigma)}$, we generated
$\widetilde{N} = 1,000$ replicate data sets,
$\widetilde{\vv{y}}_{j,(\delta,\sigma)}^1, \ldots,
\widetilde{\vv{y}}_{j,(\delta,\sigma)}^{\widetilde{N}}$, from the
same distribution as that of $\vv{y}_{j,(\delta,\sigma)}$. Again, 
this was done using a
common master batch of 1,000 standardized data sets.  
For each anchor model
$A_{j,(\delta,\sigma)}^m$, we generated $T=3,000$ Monte Carlo
samples of the model parameters $\vv{\gamma}^1,\ldots,\vv{\gamma}^T$
from the posterior distribution of $\vv{\gamma}$ conditional on
$\vv{y}_{j,(\delta,\sigma)}$.  We then estimated the ELPPD for that anchor
model fit to $\vv{y}_{j,(\delta,\sigma)}$ by the quantity
$1/\widetilde{N}\sum_{i=1}^{\widetilde{N}}\left\{
\sum_{k=1}^{10} \left[
\log\left(T^{-1}
\sum_{t=1}^T
f(\widetilde{y}_{j,(\delta,\sigma)}^{\,i,k} |
\vv{\gamma}^t)\right)\right]\right\}$, 
which provides a Monte Carlo estimate of the expected log
predictive density, $p(\tilde{\vv{y}}|\vv{y}_{j,(\delta,\sigma)})$, of a new sample,
$\tilde{\vv{y}}$, and will be large when the model has strong
predictive performance.

\begin{figure}[t!]
	\includegraphics[width=\textwidth]{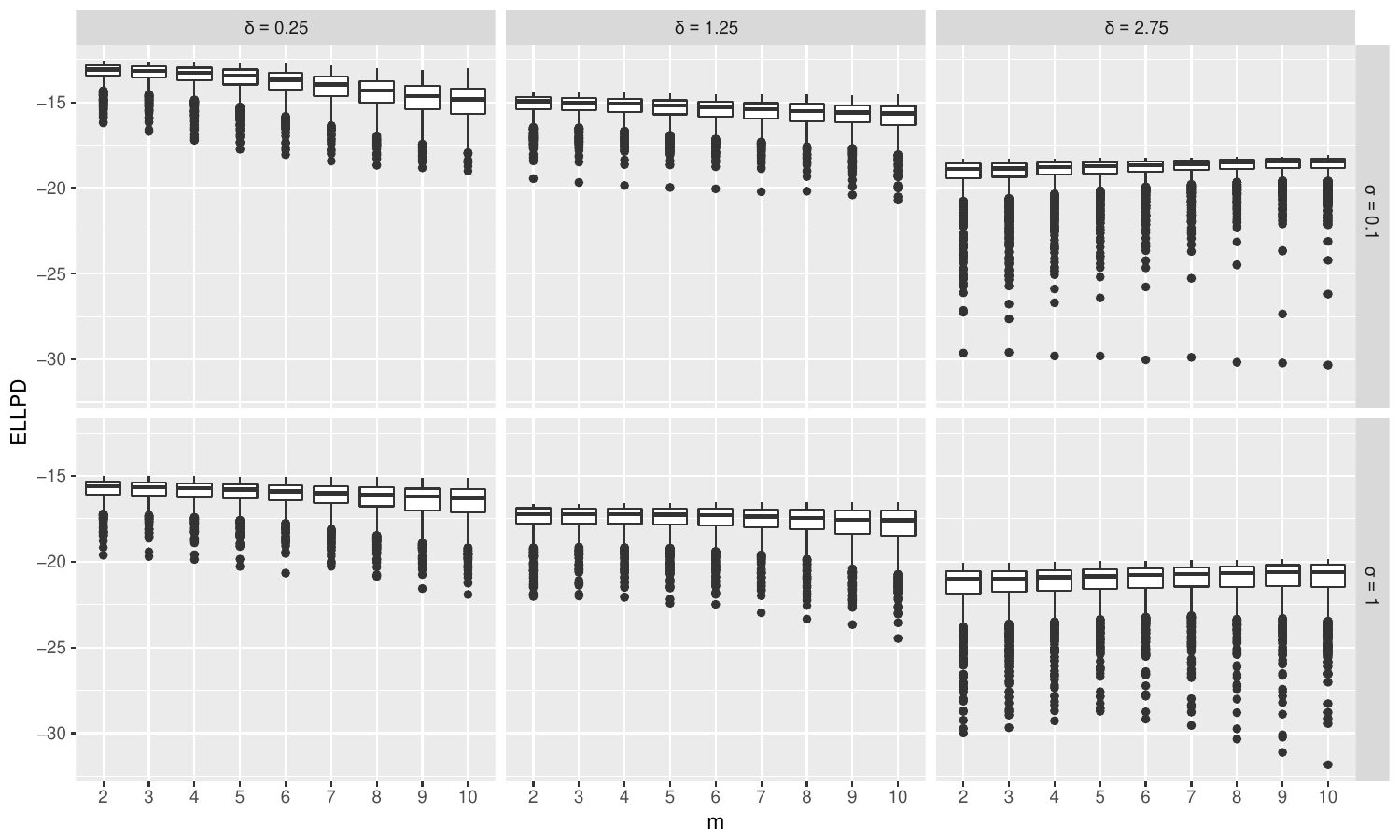}
	\caption{ \label{predictive_m} Boxplots of the values of the ELPPD of
		each simulated dataset for anchor models with $m=2,\ldots,10$.  The
		panels show results for selected experimental conditions with
		$\delta=0.25, 1.75, 2.75$ and $\sigma = 0.1, 1$. }
\end{figure}

Figure~\ref{predictive_m} shows boxplots of the 1,000 estimated ELPPD values for each
of the $(\delta, \sigma)$ simulation settings, with
$\delta=0.25, 1.75, 2.75$ and $\sigma_2 =0.10, 1,$ and each value
for the number of anchor points, $m$, between 2 and 9.  
For the settings with the largest value of $\delta=2.75$ in
which the mixture components are well-separated, there is
only a slight increase in predictive performance as the number
of anchored points increases.
For the smallest value of $\delta=0.25$, however, the
predictive performance deteriorates when the number
of anchor points is large: the ELPPD values
appear to have both lower medians and higher
variability for models with more anchor points.  For the settings in which the data were generated with
$\sigma \neq 1$, this pattern of deterioration as $m$ grows is more
apparent. 

\subsection{Simulation 2}
The second simulation sought to investigate the effect of increasing
the number of anchor points when using the anchored EM algorithm.  We
simulated
data from models~1,~2, and~3, whose parameters are given in
Table~\manuscriptrefnp{sim_models} and analyses of which are presented
in Section~\manuscriptrefnp{section:simexamples} with the same prior
and hyperparameter specification.  From each model we generated 250
data sets of size $n=100$ and fit eight different anchor models,
requiring $m_j=m$, $j=1,\ldots,k$, using the values of
$m=1,2,3,5,7,9,10,12$.  The anchor points were chosen using the
anchored EM method as well as two oracle methods.  The first set of
oracle anchor points (random oracle points) were selected randomly
from, for component~$j$, the observations generated from
component~$j$.  The second set of oracle anchor points (quantile
oracle points) were chosen to be evenly-spaced percentiles from the
true component densities, beginning at the 5th percentile and ending
at the 95th.  The model with $m=1$ uses only the 5th percentile.

\begin{figure}
	\caption{\label{fig:elppd_all} Estimated ELPPD values under the anchored EM (left), random oracle (center), and quantile oracle (right) anchor models.  Rows~1,~2 and~3 give results for models~1, ~2, and~3, respectively.}
	\includegraphics{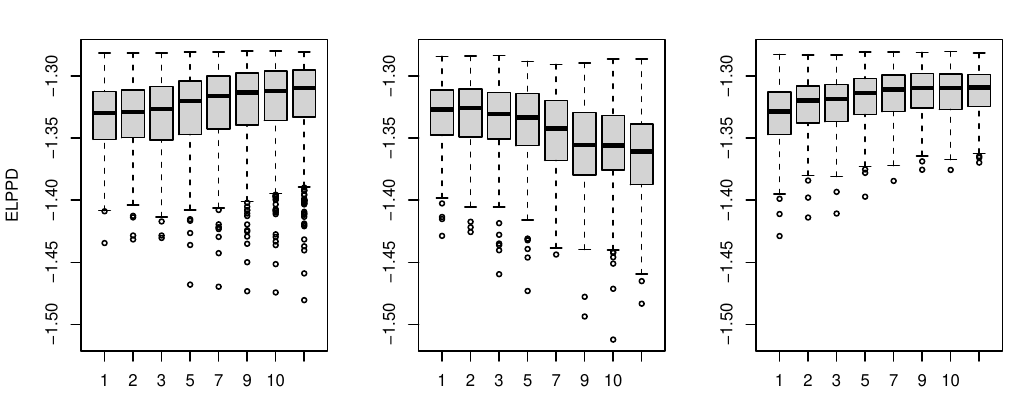}
	\includegraphics{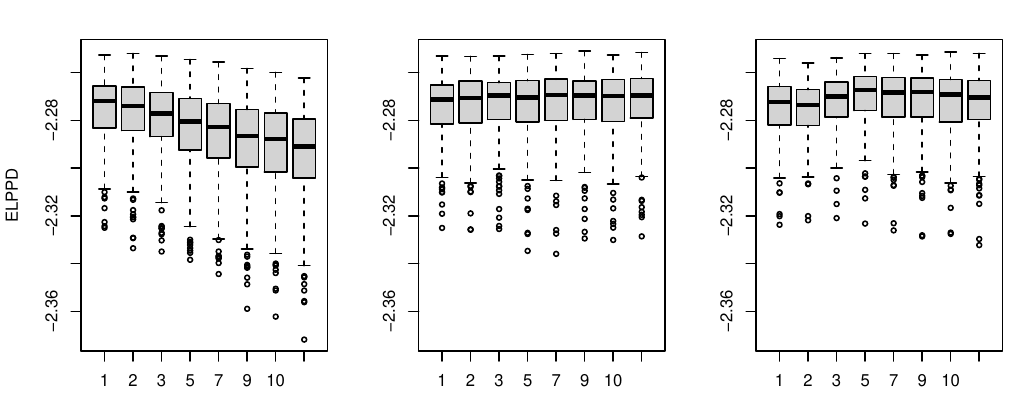}
	\includegraphics{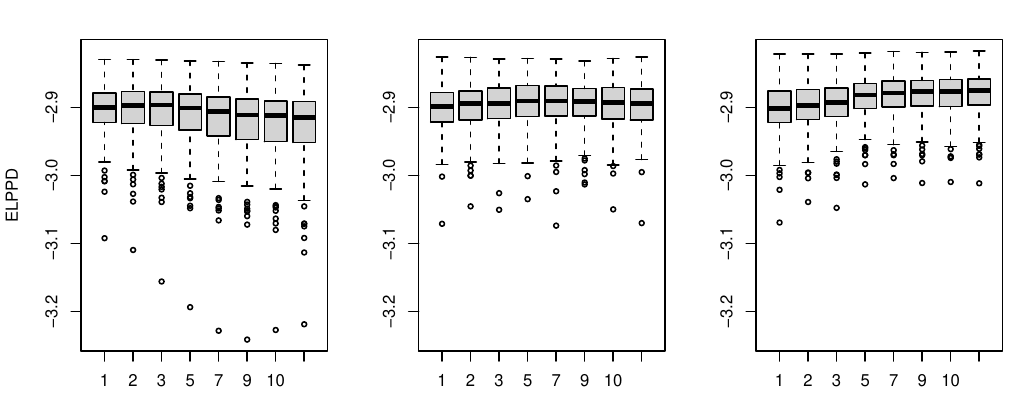}
\end{figure}

The estimated ELPPD was calculated for each data set to assess the
model's predictive performance. 
Figure~\ref{fig:elppd_all} shows boxplots of the ELPPD values for
models~1, 2, and~3 respectively.
The left-hand panels display the values under anchor models specified using anchored EM, as would occur in practice.    For model~1, the two-component scale mixture, the median predictive performance improves slightly as $m$ increases, but for model~2, the four-component location mixture, the performance deteriorates noticeably as the number of anchor points increases. For model~3 there is also a slight tendency for the ELPPD to decrease with higher values of $m$.  In all models there is more variability in the ELPPD when $m$ is large.  This pattern suggest that in models with substantial overlap among components, using many points selected by anchored EM leads to models that may introduce bias. In all models there is little apparent difference in the average ELPPD between models with $m=1$ and $m=2$.

The random and quantile oracle methods demonstrate how predictive performance changes with anchor points that represent true features of the distributions.  In the random oracle models, shown in the middle panels of Figure~\ref{fig:elppd_all} the anchor points are guaranteed to be a random sample from the component to which they are anchored.  In models~2 and~3, there is little difference in the model performance as $m$ increases.  In model~1, larger numbers of points lead to poorer predictive performance.  
The estimated ELPPD values under the quantile oracle models, are shown in the right panels of Figure~\ref{fig:elppd_all}.  As $m$ increases, these points will become increasingly representative of the features of the true component densities.  It is unsurprising that, for these methods, there is a slight increase in predictive performance as $m$ increases.

\newpage
\section{Anchored EM algorithm}
In this section we list the pseudo-code of
the Anchored EM algorithm described in
Section~\ref{section:aem} of the main
manuscript. 
% \begin{algorithm}[H]
%\caption{Anchored EM Algorithm}
%\label{A.anchored_EM Algorithm}
%\begin{spacing}{1.25}
%\begin{algorithmic}
%\item Initialize $\vv{\gamma}^0$ and select $\epsilon>0$. Set $t= 1$ and $\delta > \epsilon$.
%%\STATE 
%\WHILE{ $\delta \geq \epsilon$}
% \STATE \textbf{E step:} Calculate the unconstrained posterior
% probabilities $r_{ij}^t$ at the current value of $\vv{\gamma}^{t-1}$.
% Initialize $A^t = \emptyset$. 
%%\vskip .3cm 
%\WHILE{$|A^t| < m $}
% \STATE Find $i^\prime, j^\prime = \max_{i,j : i \not\in A^t, \; |A_j^t| < m_j} r_{ij}^t$.  %If $|A_{j^\prime}^t| < m_{j^\prime}$, 
% Set $i^\prime \in A_{j^\prime}$. 
% \STATE %While $|A_j^t| < m_j$, find $i' = \max_i r_{ij}^t, i \notinA^t$. 
%% \vskip .3cm
%  Set $\widetilde{r}_{i^\prime j^\prime}^t = 1$ and $\widetilde{r}_{i^\prime l}^t
%  = 0, l\neq j^\prime$. %,  and $i^\prime \in A_j^t$. 
%\ENDWHILE
% \vskip .2cm
%%  Do this for each $j \leq k_0$.  
%  Set $\widetilde{r}_{ij}^t = r_{ij}^t$ for $i \notin A^t,
%  j=1,\ldots,k$.  Define $q^t$ as in $M14$ using
%  $\widetilde{r}_{ij}^t$. 
%% \vskip .45cm
% \STATE \textbf{M step:}  Set $\vv{\gamma}^t =
% \argmax_{\vv{\gamma}}F(\vv{\gamma}, q^{t})$. 
%% \vskip .3cm
% \STATE $\delta = F( \vv{\gamma}^t,q^t) -
% F(\vv{\gamma}^{t-1},q^{t-1})$; t++. 
%\ENDWHILE
%\end{algorithmic}
%\end{spacing}
%\end{algorithm}

\begin{algorithm}[H]
	\caption{Anchored EM Algorithm}
	\label{A.anchored_EM Algorithm}
	\begin{spacing}{1.25}
		\begin{algorithmic}
			\item Initialize $\vv{\gamma}^0$ and select $\epsilon>0$. Set $t= 1$ and $\delta > \epsilon$.
			%\STATE 
			\WHILE{ $\delta \geq \epsilon$}
			\STATE \textbf{E step:} Calculate the unconstrained posterior
			probabilities $r_{ij}^t$ at the current value of $\vv{\gamma}^{t-1}$.
			Initialize $A^t = \emptyset$. 
			%\vskip .3cm 
			%$\WHILE{$|A^t| < m $}
			\STATE Find $A^t$ to maximize $\sum_{j=1}^{k_0}\sum_{i \in A_j}r_{ij}^t$.
			\vskip .2cm
			Set $\widetilde{r}_{ij}^t = 1$ and $\widetilde{r}_{il}^t
			= 0, l\neq j$ for  $i \in A_j^t$, $j=1,\ldots,k_0$. \\
			%  Do this for each $j \leq k_0$.  
			Set $\widetilde{r}_{ij}^t = r_{ij}^t$ for $i \notin A^t,
			j=1,\ldots,k$.  Define $q^t$ as in $M14$ using
			$\widetilde{r}_{ij}^t$. 
			% \vskip .45cm
			\STATE \textbf{M step:}  Set $\vv{\gamma}^t =
			\argmax_{\vv{\gamma}}F(\vv{\gamma}, q^{t})$. 
			% \vskip .3cm
			\STATE $\delta = F( \vv{\gamma}^t,q^t) -
			F(\vv{\gamma}^{t-1},q^{t-1})$; t++. 
			\ENDWHILE
		\end{algorithmic}
	\end{spacing}
\end{algorithm}

An exact solution to the maximization of $A^t$ above may framed as a solution to a transportation problem where $k_0$ sources must be connected to $n$ destinations and the cost of connecting source $j$ to destination $i$ is $-r_{ij}^t$.  The solution can be expressed as an $n \times k_0$ matrix $B$ of 0's and 1's, where $B_{ij}=1$ if $i \in A_j$, subject to the constraints $\sum_{j=1}^{k_{0}}B_{ij} \leq 1$ and $\sum_{i=1}^nB_{ij} = m_j$.  The function \texttt{lp.transport} in the R package \texttt{lpSolve} \citep{R_lpSolve} can be used to perform this step of the algorithm.
Alternatively, an approximate solution may be found using, for example, Vogel's approximation method and variants thereof \citep{MathMeen2004, Juman2015} or using the heuristic method described in the pseudo-code below.
\begin{algorithmic}
	\WHILE{$|A^t| < m $}
	\STATE Find $i^\prime, j^\prime = \max_{i,j : i \not\in A^t, \; |A_j^t| < m_j} r_{ij}^t$.  %If $|A_{j^\prime}^t| < m_{j^\prime}$, 
	Set $i^\prime \in A_{j^\prime}$. 
	\STATE  Set $\widetilde{r}_{i^\prime j^\prime}^t = 1$ and $\widetilde{r}_{i^\prime l}^t
	= 0, l\neq j^\prime$. %,  and $i^\prime \in A_j^t$. 
	\ENDWHILE
\end{algorithmic} \vspace{-5mm}

\section{Random permutation sampler}

%The proof of this assertion is a slight generalization of that given
%in the appendix of \cite{Fruhwirth2001} for the case of an
%exchangeable model, and in what follows, as much as possible, we try
%to combine our notational choices with those in~\cite{Fruhwirth2001}.

As noted in Section~\ref{section:sampling}, to prove that the random
permutation sampler is a Gibbs sampler that
generates draws from the target posterior
distribution we need to show that the accepted randomly permuted draw
$(\vv{\gamma},\vv{\eta})$ is in fact a draw from the distribution of
$(\vv{\gamma},\vv{\eta})$ given $(\vv{s}, \vv{y})$.  Let $C$ be a
measurable subset and let the symbol $p_T(\cdot | \cdot )$
denote generically a conditional density under the target posterior
for the given 
model (either the exchangeable or the anchor model).
We note first that the
absolute values of the Jacobian determinants for the $k!$ permutations
and their inverses are all equal to one and can therefore be ignored
in the following derivation. 
Then we have
\begin{align} 
P((\vv{\gamma},\vv{\eta}) \in C| \vv{s}, \vv{y}) \nonumber
& =
\sum_{q = 1}^{k!} \int_{\rho_q^{-1}(C)}
\frac{p_T(\rho_q(\rho_q^{-1}( \vv{\gamma},\vv{\eta}   ) ) | \vv{y})} 
{\sum_{h=1}^{k!}
	p_T(\rho_h(\rho_q^{-1}( \vv{\gamma},\vv{\eta})) | \vv{y})
} 
p_T(\rho_q^{-1}(   \vv{\gamma},\vv{\eta} )| \vv{s}, \vv{y}) \,
d(\rho_q^{-1}(   \vv{\gamma},\vv{\eta} ))  \\   \nonumber
& =
\sum_{q = 1}^{k!} \int_C
\frac{p_T(\rho_q( \vv{\gamma},\vv{\eta}  ) | \vv{y})} 
{\sum_{h=1}^{k!}
	p_T(\rho_h(\vv{\gamma},\vv{\eta}) | \vv{y})
} 
p_T( \vv{\gamma},\vv{\eta} | \vv{s}, \vv{y}) \,
d(  \vv{\gamma},\vv{\eta} )  \\   \nonumber
& =
\int_C 
\frac{\sum_{q = 1}^{k!}  p_T(\rho_q( \vv{\gamma},\vv{\eta}  ) | \vv{y})} 
{\sum_{h=1}^{k!}
	p_T(\rho_h(\vv{\gamma},\vv{\eta}) | \vv{y})
} 
p_T( \vv{\gamma},\vv{\eta} | \vv{s}, \vv{y}) \,
d(  \vv{\gamma},\vv{\eta} )  \\   \nonumber
& =
\int_C 
p_T( \vv{\gamma},\vv{\eta} | \vv{s}, \vv{y}) \,
d(  \vv{\gamma},\vv{\eta} )  \\   \nonumber
& =
P_T((\vv{\gamma},\vv{\eta}) \in C| \vv{s}, \vv{y}), \nonumber
\end{align}
proving the claim.
Note that for the exchangeable model
all relabelings will
have the same probability $1/k!$ of being selected and the algorithm
reduces to the original algorithm proposed in ~\cite{Fruhwirth2001}.

\section{Univariate examples with $m_j=2$}
In this section, we revisit the univariate examples of Section~\manuscriptref{section::Examples} using two anchor points per component.  The oracle anchor points when $m_j=2$ are chosen as the 5th and 95th percentiles of the true component distributions.  There is no minimum-entropy method demonstrated for this case.

\begin{figure}
	\includegraphics{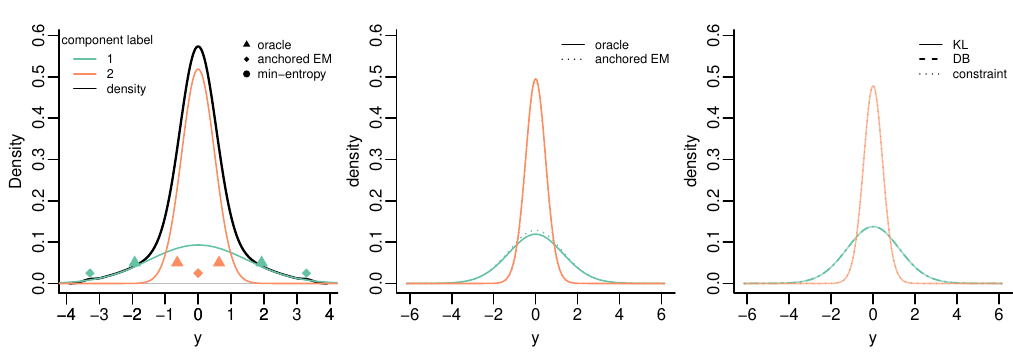}
	\caption{\label{fig:a_model1} Left panel: anchor points and true scaled component densities for model~1. Middle panel: estimated scaled component densities for the anchor models.  Right panel: estimated scaled component densities for the relabeling methods and prior constraints.}
\end{figure}

\begin{figure}
	\includegraphics{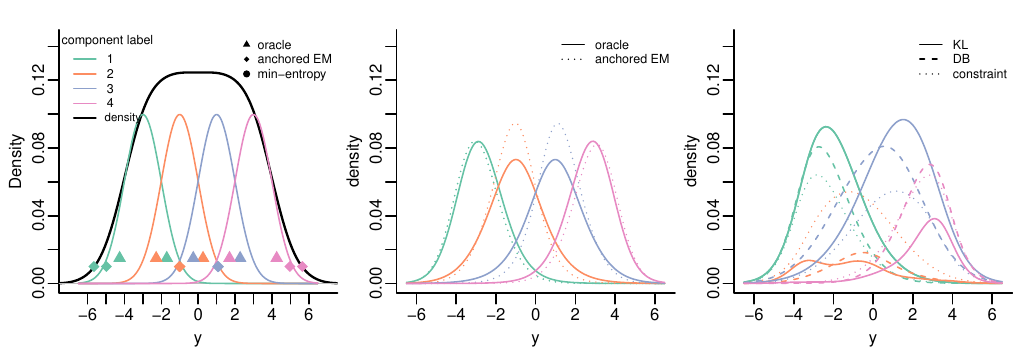}
	\caption{\label{fig:a_model2} Left panel: anchor points and true scaled component densities for model~2. Middle panel: estimated scaled component densities for the anchor models.  Right panel: estimated scaled component densities for the relabeling methods and prior constraints.}
\end{figure}

\begin{figure}
	\includegraphics{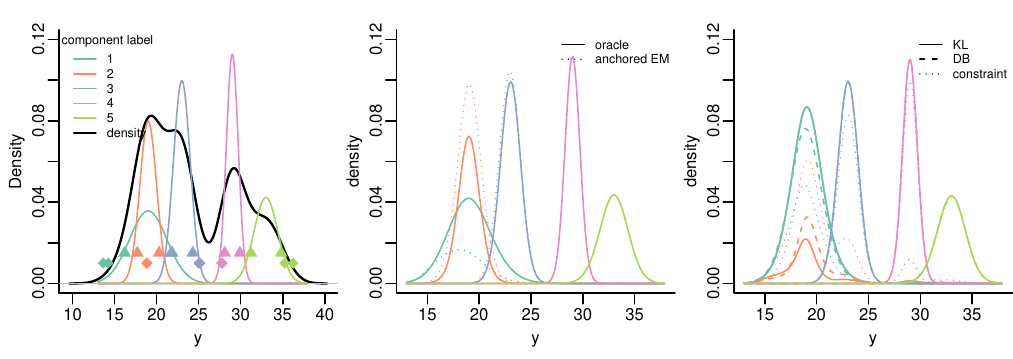}
	\caption{\label{fig:a_model3} Left panel: anchor points and true scaled component densities for model~3. Middle panel: estimated scaled component densities for the anchor models.  Right panel: estimated scaled component densities for the relabeling methods and prior constraints.}
\end{figure}

Figure~\ref{fig:a_model1} shows the scaled component densities under Model~1 for the anchor models and relabeling methods.  Table~\ref{table:a_model1} gives posterior means of the model parameters for all methods, including the anchor models with $m_j=1$ as presented in the main manuscript.
% latex table generated in R 4.0.0 by xtable 1.8-4 package
% Tue Jun 30 10:23:00 2020
\begin{table}[ht]
	\caption{\label{table:a_model1} Posterior means of the component-specific parameters for Model 1.}
	\centering
	\small
	\begin{tabular}{rllll}
		\hline
		& $\theta_1$ & $\theta_2$ & error & time (s) \\ 
		\hline
		true & 0 & 0 & 0 & 0 \\ 
		oracle anchors ($m_j=2$)& 0.0004 (0.0014) & -0.0008 (0.0005) & 0.0012 & N/A \\ 
		anchored EM ($m_j=2$) & -0.0004 (0.0014) & -0.0007 (0.0005) & 0.0011 & 1.03 \\ 
		oracle anchors ($m_j=1$) & 0.0010 (0.0010) & -0.0010 (0.0010) & 0.0020 & N/A \\ 
		anchored EM ($m_j=1$) & -0.0027 (0.0014) & -0.0009 (0.0005) & 0.0036 & 1.03 \\ 
		min-entropy($m_j=1$)  & -0.0003 (0.0014) & -0.0006 (0.0005) & 0.0009 & 0.50 \\ 
		KL & 0.0005 (0.0013) & -0.0008 (0.0005) & 0.0013 & 1102.78 \\ 
		DB & 0.0005 (0.0013) & -0.0008 (0.0005) & 0.0013 & 547.59 \\ 
		constraint & 0.0005 (0.0013) & -0.0008 (0.0005) & 0.0013 & N/A \\  
		\hline
		\hline
		& $\sigma_1$ & $\sigma_2$ & error &  \\ 
		\hline
		true &  1.5 &  0.5 & 0 &\\ 
		oracle anchors ($m_j=2$)& 1.356 (0.0013) & 0.485 (0.0005) & 0.1593 &  \\ 
		anchored EM ($m_j=2$)& 1.333 (0.0013) & 0.470 (0.0006) & 0.1973 &  \\ 
		oracle anchors ($m_j=1$) & 0.867 (0.0035) & 0.866 (0.0035) & 0.9988 &  \\ 
		anchored EM ($m_j=1$)& 1.320 (0.0013) & 0.467 (0.0006) & 0.2125 &  \\ 
		min-entropy ($m_j=1$) & 1.320 (0.0013) & 0.468 (0.0006) & 0.2122 &  \\ 
		KL & 1.307 (0.0013) & 0.464 (0.0005) & 0.2293 &  \\ 
		DB & 1.307 (0.0013) & 0.464 (0.0006) & 0.2293 &  \\ 
		constraint & 1.733 (0.0013) & 0.220 (0.0005) & 0.5129 &  \\ 
		\hline
		\hline
		& $\eta_1$ & $\eta_2$ &error &  \\ 
		\hline
		true & 0.35 & 0.65 & 0 &\\ 
		oracle anchors ($m_j=2$) & 0.395 (0.0008) & 0.605 (0.0008) & 0.0907 &  \\ 
		anchored EM ($m_j=2$)& 0.418 (0.0008) & 0.582 (0.0008) & 0.1363 &  \\ 
		oracle anchors ($m_j=1$) & 0.501 (0.0009) & 0.499 (0.0009) & 0.3020 &  \\ 
		anchored EM ($m_j=1$) & 0.430 (0.0009) & 0.570 (0.0009) & 0.1594 &  \\ 
		min-entropy ($m_j=1$)& 0.428 (0.0009) & 0.572 (0.0009) & 0.1568 &  \\ 
		KL & 0.439 (0.0009) & 0.561 (0.0009) & 0.1781 &  \\ 
		DB & 0.439 (0.0009) & 0.561 (0.0009) & 0.1782 &  \\ 
		constraint & 0.439 (0.0009) & 0.561 (0.0009) & 0.1781 &  \\ 
		\hline
	\end{tabular}
\end{table}
Figure~\ref{fig:a_model2} shows the scaled component densities under Model~2 and  Table~\ref{table:a_model2} gives posterior means of the model parameters for all methods, including the anchor models with $m_j=1$ as presented in the main manuscript.
The results for Model~3 are shown in Figure~\ref{fig:a_model3} and Table~\ref{table:a_model3}.   

In Model~1, the anchor models with two anchor points provide parameter estimates with higher accuracy than the ones with one anchor points.  
This improvement is, however, not seen for all parameters in Model~2 and Model~3.  The anchored EM model with $m_j=1$ provides more accurate estimates of both $\vv{\theta}$ and $\vv{\eta}$ for Model~2, although $m_j=2$ produces better estimates of $\vv{\sigma}$.  For Model~3, the anchored EM model is more accurate with $m_j=1$ than with $m_j=2$.
% latex table generated in R 4.0.0 by xtable 1.8-4 package
% Tue Jun 30 10:55:01 2020
\begin{table}[ht]
	\caption{\label{table:a_model2} Posterior means of the component-specific parameters for Model 2.}
	\centering
	\small
	\resizebox{\textwidth}{!}{	\begin{tabular}{rllllll}
			\hline
			& $\theta_1$ & $\theta_2$ & $\theta_3$ & $\theta_4$ & error & time (s) \\ 
			\hline
			true & -3 & -1 & 1 & 3 & 0 &\\ 
			oracle anchors ($m_j=2$) & -2.802 (0.0037) & -0.978 (0.0059) & 0.979 (0.0057) & 2.800 (0.0038) & 0.442 & N/A \\ 
			anchored EM ($m_j=2$) & -3.104 (0.0051) & -1.061 (0.0050) & 1.138 (0.0050) & 3.140 (0.0051) & 0.443 & 21.39 \\
			oracle anchors ($m_j=1$) & -2.819 (0.0049) & -0.960 (0.0072) & 0.956 (0.0072) & 2.818 (0.0049) & 0.448 & N/A \\ 
			anchored EM  ($m_j=1$) & -2.983 (0.0069) & -0.960 (0.0073) & 0.997 (0.0073) & 2.997 (0.0068) & 0.063 & 24.87 \\ 
			min-entropy ($m_j=1$)& -2.967 (0.0116) & -1.653 (0.0095) & 1.965 (0.0088) & 2.864 (0.0134) & 1.786 & 10.37 \\ 
			KL & -2.208 (0.0078) & -0.942 (0.0376) & 1.288 (0.0090) & 1.864 (0.0318) & 2.274 & 6134.31 \\ 
			DB & -2.536 (0.0141) & -0.259 (0.0393) & 0.278 (0.0136) & 2.520 (0.0199) & 2.408 & 159.00 \\ 
			constraint & -3.477 (0.0230) & -1.039 (0.0129) & 1.027 (0.0127) & 3.492 (0.0225) & 1.035 & N/A \\ 
			\hline
			\hline
			& $\sigma_1$ & $\sigma_2$ & $\sigma_3$ & $\sigma_4$ & error & \\ 
			\hline
			true &    1 &    1 &    1 &    1 & 0  & \\ 
			oracle anchors ($m_j=2$) & 1.109 (0.0020) & 1.257 (0.0030) & 1.259 (0.0031) & 1.108 (0.0020) & 0.7333 &  \\ 
			anchored EM ($m_j=2$) & 1.011 (0.0020) & 0.954 (0.0027) & 0.945 (0.0026) & 1.001 (0.0020) & 0.1135 &  \\
			oracle anchors ($m_j=1$) & 1.024 (0.0026) & 1.067 (0.0041) & 1.063 (0.0039) & 1.026 (0.0025) & 0.1805 &  \\ 
			anchored EM ($m_j=1$)& 1.074 (0.0025) & 0.984 (0.0031) & 0.984 (0.0032) & 1.069 (0.0025) & 0.1752 &  \\ 
			min-entropy ($m_j=1$)& 1.121 (0.0040) & 1.089 (0.0036) & 1.042 (0.0034) & 1.168 (0.0043) & 0.4192 &  \\ 
			KL & 1.243 (0.0037) & 0.699 (0.0069) & 1.497 (0.0023) & 0.766 (0.0068) & 1.270 &  \\ 
			DB & 1.071 (0.0033) & 0.687 (0.0028) & 1.476 (0.0050) & 0.970 (0.0031) & 0.8893 &  \\ 
			constraint & 0.981 (0.0038) & 1.123 (0.0048) & 1.122 (0.0048) & 0.979 (0.0038) & 0.2848 &  \\  
			\hline
			\hline
			& $\eta_1$ & $\eta_2$ & $\eta_3$ & $\eta_4$  & error & \\ 
			\hline
			true & 0.25 & 0.25 & 0.25 & 0.25 & 0 &\\ 
			oracle anchors ($m_j=2$) & 0.245 (0.0007) & 0.254 (0.0009) & 0.255 (0.0009) & 0.246 (0.0007) & 0.0181 &  \\ 
			anchored EM ($m_j=2$) & 0.236 (0.0010) & 0.269 (0.0010) & 0.266 (0.0010) & 0.229 (0.0009) & 0.0702 &  \\ 
			oracle anchors ($m_j=1$) & 0.247 (0.0011) & 0.252 (0.0013) & 0.251 (0.0013) & 0.250 (0.0010) & 0.0059 &  \\ 
			anchored EM ($m_j=1$)& 0.255 (0.0012) & 0.247 (0.0012) & 0.247 (0.0012) & 0.251 (0.0012) & 0.0124 &  \\ 
			min-entropy ($m_j=1$) & 0.222 (0.0016) & 0.283 (0.0016) & 0.263 (0.0014) & 0.232 (0.0018) & 0.0923 &  \\ 
			KL & 0.353 (0.0015) & 0.075 (0.0006) & 0.459 (0.0017) & 0.114 (0.0008) & 0.6230 &  \\ 
			DB & 0.273 (0.0015) & 0.077 (0.0007) & 0.427 (0.0022) & 0.223 (0.0014) & 0.4010 &  \\ 
			constraint & 0.224 (0.0017) & 0.276 (0.0020) & 0.277 (0.0019) & 0.224 (0.0017) & 0.1047 &  \\  
			\hline
	\end{tabular} }
\end{table}

% latex table generated in R 4.0.0 by xtable 1.8-4 package
% Tue Jun 30 12:07:41 2020
\begin{table}[ht]
	\caption{\label{table:a_model3} Posterior means of the component-specific parameters for Model 3.}
	\centering
	\resizebox{\textwidth}{!}{	\begin{tabular}{rlllllll}
			\hline
			& $\theta_1$ & $\theta_2$ & $\theta_3$ &$\theta_4$ & $\theta_5$ & error & time (s) \\ 
			\hline
			true & 19 & 19 & 23 & 29 & 33 & 0 &\\ 
			oracle anchors ($m_j=2$) & 18.990 (0.0053) & 19.012 (0.0020) & 23.023 (0.0011) & 29.007 (0.0006) & 32.993 (0.0016) & 0.059 & N/A \\ 
			anchored EM ($m_j=2$) & 16.822 (0.0224) & 19.044 (0.0015) & 22.954 (0.0012) & 29.006 (0.0006) & 32.990 (0.0016) & 2.284 & 16.11 \\ 
			oracle anchors ($m_j=1$) & 18.955 (0.0049) & 18.958 (0.0049) & 23.030 (0.0014) & 29.006 (0.0006) & 32.990 (0.0016) & 0.135 & N/A \\ 
			anchored EM ($m_j=1$)& 17.273 (0.0290) & 19.049 (0.0019) & 22.975 (0.0014) & 29.006 (0.0006) & 32.991 (0.0016) & 1.816 & 16.60 \\ 
			min-entropy ($m_j=1$) & 18.948 (0.0014) & 23.011 (0.0010) & 29.006 (0.0006) & 33.031 (0.0056) & 33.838 (0.0100) & 14.94 & 10.56 \\ 
			KL & 19.086 (0.0026) & 20.087 (0.0868) & 23.028 (0.0016) & 29.005 (0.0007) & 32.992 (0.0018) & 1.215 & 986.31 \\ 
			DB & 18.799 (0.0062) & 20.335 (0.0831) & 23.065 (0.0044) & 29.007 (0.0007) & 32.994 (0.0018) & 1.613 & 285.12 \\ 
			constraint & 17.848 (0.0255) & 20.062 (0.0287) & 23.856 (0.0351) & 29.266 (0.0117) & 33.167 (0.0150) & 3.504 & N/A \\
			\hline
			\hline
			& $\sigma_1$ & $\sigma_2$ & $\sigma_3$ &$\sigma_4$ & $\sigma_5$ & error &  \\ 
			\hline
			true & 2.236 &    1 &    1 & 0.7071 & 1.414 & 0 &\\ 
			oracle anchors ($m_j=2$) & 2.059 (0.0028) & 1.018 (0.0020) & 0.973 (0.0008) & 0.717 (0.0005) & 1.363 (0.0012) & 0.284 &  \\ 
			anchored EM ($m_j=2$) & 1.377 (0.0067) & 1.125 (0.0031) & 1.023 (0.0009) & 0.719 (0.0005) & 1.364 (0.0012) & 1.070 &  \\ 
			oracle anchors $(m_j=1)$ & 1.371 (0.0048) & 1.380 (0.0048) & 0.973 (0.0010) & 0.719 (0.0005) & 1.364 (0.0012) & 1.334 &  \\ 
			anchored EM  $(m_j=1)$& 1.490 (0.0081) & 1.073 (0.0042) & 1.011 (0.0009) & 0.719 (0.0005) & 1.363 (0.0012) & 0.892&  \\ 
			min-entropy $(m_j=1)$ & 1.606 (0.0010) & 1.011 (0.0007) & 0.718 (0.0005) & 1.067 (0.0025) & 1.134 (0.0037) & 1.563 &  \\ 
			KL & 1.581 (0.0023) & 0.986 (0.0062) & 0.973 (0.0013) & 0.716 (0.0011) & 1.358 (0.0007) & 0.760 &  \\ 
			DB & 1.579 (0.0036) & 0.989 (0.0052) & 0.971 (0.0012) & 0.716 (0.0006) & 1.358 (0.0013) & 0.761 &  \\ 
			constraint & 1.322 (0.0049) & 1.273 (0.0060) & 0.934 (0.0024) & 0.746 (0.0018) & 1.338 (0.0019) & 1.368 &  \\
			\hline
			\hline
			& $\eta_1$ & $\eta_2$ & $\eta_3$ &$\eta_4$ & $\eta_5$ & error &  \\ 
			\hline
			true & 0.2 & 0.2 & 0.25 & 0.2 & 0.15 & 0 &\\ 
			oracle anchors ($m_j=2$) & 0.219 (0.0006) & 0.187 (   NA) & 0.244 (0.0003) & 0.201 (0.0001) & 0.149 (0.0001) & 0.0405 &  \\ 
			anchored EM ($m_j=2$) & 0.094 (0.0015) & 0.286 (0.0013) & 0.269 (0.0003) & 0.201 (0.0001) & 0.149 (0.0001) & 0.2134 &  \\ 
			oracle anchors $(m_j=1)$ & 0.201 (0.0008) & 0.204 (0.0008) & 0.245 (0.0003) & 0.201 (0.0001) & 0.149 (0.0001) & 0.0127 &  \\ 
			anchored EM $(m_j=1)$ & 0.132 (0.0021) & 0.255 (0.0019) & 0.264 (0.0003) & 0.201 (0.0001) & 0.149 (0.0001) & 0.1387 &  \\ 
			min-entropy $(m_j=1)$ & 0.392 (0.0002) & 0.255 (0.0002) & 0.200 (0.0002) & 0.088 (0.0004) & 0.064 (0.0004) & 0.4954 &  \\ 
			KL & 0.334 (0.0011) & 0.074 (0.0008) & 0.246 (0.0005) & 0.198 (0.0002) & 0.147 (0.0002) & 0.2688 &  \\ 
			DB & 0.303 (0.0016) & 0.106 (0.0013) & 0.245 (0.0005) & 0.198 (0.0002) & 0.147 (0.0002) & 0.2063 &  \\ 
			constraint & 0.189 (0.0023) & 0.253 (0.0012) & 0.228 (0.0010) & 0.186 (0.0007) & 0.143 (0.0003) & 0.1060 &  \\  
			\hline
	\end{tabular} }
\end{table}

\clearpage
\section{Analysis of the  SisFall data}
This section provides additional details on the data analysis example presented in Section~\ref{section::Falldata} of the main manuscript.

\paragraph{Data.} The full SisFall data set collected by \cite{Sucetal2017}, with additional details on the experimental procedure, is available at \url{http://sistemic.udea.edu.co/en/research/projects/english-falls/}.  

\paragraph{Additional tables.} 
Table~\ref{table::fall_thetaj} gives posterior means of $\vv{\theta}_j$ for each mixture component. Table~\ref{table::fall_allocations} gives the full table of posterior allocation probabilities for each of the activities considered in the analysis.  Activities beginning with ``D'' are ADLs and activities beginning with ``F'' are falls.

% latex table generated in R 3.5.2 by xtable 1.8-3 package
% Fri May 22 14:11:13 2020
\begin{table}[H]
	\centering
	\caption{\label{table::fall_thetaj} Posterior means of
		$\vv{\theta}_j$ for each component from the SisFall data. } 
	\begin{tabularx}{\textwidth}{p{5cm}XXXXX}
		\hline
		Component & 1  & 2  & 3  & 4  & 5\\ 
		\hline
		$\log(\max_tSMV_t)$ & 5.769 & 7.304 & 7.169 & 6.972 & 6.366 \\ 
		$\log(\min_tSMV_t)$ & 5.323 & 3.661 & 1.686 & 4.626 & 3.795 \\ 
		$\log(\max_t|SMV_t-SMV_{t-1}|)$ & 3.162 & 5.963 & 5.802 & 5.021 & 3.979 \\ 
		\hline
	\end{tabularx}
\end{table}

% latex table generated in R 3.5.2 by xtable 1.8-3 package
% Fri May 15 16:36:10 2020
\begin{table}[ht]
	\small
	\centering
	\caption{\label{table::fall_allocations} Posterior allocation probabilities for selected activities in the SisFall dataset.}
	\begin{tabular}{p{11cm}rrrrr}
		\hline
		Activity &  \multicolumn{5}{l}{Component }\\
		& 1 & 2 & 3 & 4 & 5 \\ 
		\hline
		D01 Walking slowly & 0.099 & 0.008 & 0.000 & 0.490 & 0.403 \\ 
		D02 Walking quickly & 0.000 & 0.086 & 0.000 & 0.066 & 0.848 \\ 
		D03 Jogging slowly & 0.000 & 0.000 & 1.000 & 0.000 & 0.000 \\ 
		D04 Jogging quickly & 0.000 & 0.011 & 0.989 & 0.000 & 0.000 \\ 
		D05 Walking upstairs and downstairs slowly & 0.059 & 0.027 & 0.000 & 0.887 & 0.027 \\ 
		D06 Walking upstairs and downstairs quickly & 0.000 & 0.796 & 0.188 & 0.000 & 0.016 \\ 
		D07 Slowly sit in a half height chair, wait a moment, and up slowly  & 0.981 & 0.000 & 0.000 & 0.006 & 0.013 \\ 
		D08 Quickly sit in a half height chair, wait a moment, and up quickly & 0.000 & 0.019 & 0.001 & 0.024 & 0.956 \\ 
		D09 Slowly sit in a low height chair, wait a moment, and up slowly & 0.245 & 0.050 & 0.000 & 0.575 & 0.131 \\ 
		D10 Quickly sit in a low height chair, wait a moment, and up quickly & 0.001 & 0.042 & 0.001 & 0.148 & 0.808 \\ 
		D11 Sitting a moment, trying to get up, and collapse into a chair & 0.000 & 0.541 & 0.003 & 0.454 & 0.002 \\ 
		D12 Sitting a moment, lying slowly, wait a moment, and sit again & 0.938 & 0.020 & 0.000 & 0.035 & 0.007 \\ 
		D13 Sitting a moment, lying quickly, wait a moment, and sit again & 0.173 & 0.110 & 0.000 & 0.695 & 0.022 \\ 
		D14 Being on one's back change to lateral position, wait a moment, and change to one's back & 0.942 & 0.001 & 0.000 & 0.038 & 0.019 \\ 
		D15 Standing, slowly bending at knees, and getting up & 0.978 & 0.000 & 0.000 & 0.010 & 0.012 \\ 
		D16 Standing, slowly bending without bending knees, and getting up & 0.986 & 0.000 & 0.000 & 0.008 & 0.005 \\ 
		D17 Standing, get into a car, remain seated and get out of the car & 0.797 & 0.003 & 0.000 & 0.156 & 0.045 \\ 
		D18 Stumble while walking & 0.000 & 0.957 & 0.003 & 0.040 & 0.000 \\ 
		D19 Gently jump without falling (trying to reach a high object) & 0.000 & 0.308 & 0.033 & 0.000 & 0.658 \\ 
		F01 Fall forward while walking caused by a slip & 0.000 & 0.915 & 0.001 & 0.084 & 0.000 \\ 
		F02 Fall backward while walking caused by a slip & 0.000 & 0.302 & 0.001 & 0.697 & 0.000 \\ 
		F03 Lateral fall while walking caused by a slip & 0.000 & 0.250 & 0.000 & 0.750 & 0.000 \\ 
		F04 Fall forward while walking caused by a trip & 0.000 & 0.938 & 0.001 & 0.061 & 0.000 \\ 
		F05 Fall forward while jogging caused by a trip & 0.000 & 0.799 & 0.201 & 0.000 & 0.000 \\ 
		F06 Vertical fall while walking caused by fainting & 0.000 & 0.058 & 0.000 & 0.942 & 0.000 \\ 
		F07 Fall while walking, with use of hands in a table to dampen fall, caused by fainting & 0.000 & 0.211 & 0.001 & 0.788 & 0.000 \\ 
		F08 Fall forward when trying to get up & 0.001 & 0.586 & 0.001 & 0.412 & 0.000 \\ 
		F09 Lateral fall when trying to get up & 0.000 & 0.074 & 0.000 & 0.926 & 0.000 \\ 
		F10 Fall forward when trying to sit down & 0.000 & 0.609 & 0.001 & 0.390 & 0.000 \\ 
		F11 Fall backward when trying to sit down & 0.000 & 0.230 & 0.002 & 0.760 & 0.008 \\ 
		F12 Lateral fall when trying to sit down & 0.000 & 0.239 & 0.001 & 0.759 & 0.000 \\ 
		F13 Fall forward while sitting, caused by fainting or falling asleep & 0.014 & 0.726 & 0.001 & 0.124 & 0.135 \\ 
		F14 Fall backward while sitting, caused by fainting or falling asleep & 0.001 & 0.527 & 0.003 & 0.467 & 0.002 \\ 
		F15 Lateral fall while sitting, caused by fainting or falling asleep & 0.000 & 0.439 & 0.001 & 0.561 & 0.000 \\ 
		\hline
	\end{tabular}
\end{table}

\end{document}